\pdfoutput=1
\documentclass[twocolumn,twocolappendix]{aastex631}
\usepackage{here}
\usepackage{url}
\usepackage{color,ulem}

\defcitealias{2021ApJ...906L...3T}{T+21}
\shorttitle{Circumstellar structure of Tycho's SNR--dynamics}
\shortauthors{Kobashi et al.}
\graphicspath{{./}{figures/}}
\begin{document}

\title{Exploring the circumstellar environment of Tycho's supernova remnant--I. The hydrodynamic evolution of the shock}

\author[0009-0008-4215-1049]{Ryosuke Kobashi}
\affiliation{Department of Astronomy, Kyoto University, Kitashirakawa, Oiwake-cho, Sakyo-ku, Kyoto 606-8502, Japan}

\author[0000-0002-2899-4241]{Shiu-Hang Lee}
\affiliation{Department of Astronomy, Kyoto University, Kitashirakawa, Oiwake-cho, Sakyo-ku, Kyoto 606-8502, Japan}
\affiliation{Kavli Institute for the Physics and Mathematics of the Universe (WPI), The University of Tokyo, Kashiwa 277-8583, Japan}

\author[0000-0002-4383-0368]{Takaaki Tanaka}
\affiliation{Department of Physics, Konan University, 8-9-1 Okamoto, Higashinada, Kobe, Hyogo 658-8501, Japan}

\author[0000-0003-2611-7269]{Keiichi Maeda}
\affiliation{Department of Astronomy, Kyoto University, Kitashirakawa, Oiwake-cho, Sakyo-ku, Kyoto 606-8502, Japan}

\correspondingauthor{Ryosuke Kobashi}
\email{kobashi@kusastro.kyoto-u.ac.jp}

%% Note that the \and command from previous versions of AASTeX is now
%% depreciated in this version as it is no longer necessary. AASTeX 
%% automatically takes care of all commas and "and"s between authors names.

%% AASTeX 6.31 has the new \collaboration and \nocollaboration commands to
%% provide the collaboration status of a group of authors. These commands 
%% can be used either before or after the list of corresponding authors. The
%% argument for \collaboration is the collaboration identifier. Authors are
%% encouraged to surround collaboration identifiers with ()s. The 
%% \nocollaboration command takes no argument and exists to indicate that
%% the nearby authors are not part of surrounding collaborations.

%% Mark off the abstract in the ``abstract'' environment. 
\begin{abstract}
Among Type Ia supernova remnants (SNRs), Tycho's SNR has been considered as a typical object from the viewpoints of its spectroscopic, morphological and environmental properties. A recent reanalysis of Chandra data shows that its forward shock is experiencing a substantial deceleration since around 2007, which suggests recent shock interactions with a dense medium as a consequence of the cavity-wall environment inside a molecular cloud. Such a non-uniform environment can be linked back to the nature and activities of its progenitor. %From these information, knowing Tycho's age, we can strictly give a constraint on Tycho's environment and thus progenitor.
In this study, we perform hydrodynamic simulations to characterize Tycho's cavity-wall environment using the latest multi-epoch proper motion measurements of the forward shock. A range of parameters for the environment is explored in the hydrodynamic models to fit with the observation data for each azimuthal region. 
%density in cavity and distance (to the earth) which is necessary to convert proper motions to physical values. Using the correspondent values, we calculate the evolution of Tycho's shock using %, and obtain the picture of Tycho's CSM. 
%The calibrated distance at 2003 is $\sim3$~kpc (uniform cavity case) and $\sim3.5$~kpc (wind-cavity case). 
Our results show that a wind-like cavity with $\rho(r)\propto r^{-2}$ reconciles with the latest data better than a uniform medium with a constant density. 
%with $\rho(r)=\rho_0$ 
%in terms of reproducing the fast proper motion. 
In addition, our best-fit model favors an anisotropic wind with an azimuthally varying wind parameter. The overall result indicates a mass-loss rate which is unusually high for the conventional single-degenerate explosion scenario.
%, assuming a low wind velocity. 
%The azimuthal variation of the estimated cloud density can be attributed to the clumpiness of the cloud. 
\end{abstract}

%% Keywords should appear after the \end{abstract} command. 
%% The AAS Journals now uses Unified Astronomy Thesaurus concepts:
%% https://astrothesaurus.org
%% You will be asked to selected these concepts during the submission process
%% but this old "keyword" functionality is maintained in case authors want
%% to include these concepts in their preprints.


%% From the front matter, we move on to the body of the paper.
%% Sections are demarcated by \section and \subsection, respectively.
%% Observe the use of the LaTeX \label
%% command after the \subsection to give a symbolic KEY to the
%% subsection for cross-referencing in a \ref command.
%% You can use LaTeX's \ref and \label commands to keep track of
%% cross-references to sections, equations, tables, and figures.
%% That way, if you change the order of any elements, LaTeX will
%% automatically renumber them.
%%
%% We recommend that authors also use the natbib \citep
%% and \citet commands to identify citations.  The citations are
%% tied to the reference list via symbolic KEYs. The KEY corresponds
%% to the KEY in the \bibitem in the reference list below. 

\section{Introduction} \label{sec:intro}

Despite an incomplete understanding of their progenitor systems and populations, Type Ia supernovae (SNe) originating from the thermonuclear explosions of white dwarfs in binary systems have been used as a standard candle for cosmological studies \citep{1998AJ....116.1009R,1999ApJ...517..565P}. The nature of the progenitor system of SNe Ia is a matter of hot debate, which is now mainly split into two camps of scenarios depending on whether the companion star is another white dwarf or a non-degenerate star, which are nowadays referred to as the double degenerate (DD) and single degenerate (SD) scenario, respectively. 

Tycho's supernova remnant (Tycho's SNR, SN1572, G120.1+1.4; hereafter Tycho) is considered to originate in a normal Type Ia SN explosion that happened in November of year 1572. A wealth of observations have been done, such as spectroscopic observations and modeling of the thermal X-ray emission from the ejecta  \citep[e.g.,][]{2006ApJ...645.1373B,2014ApJ...780..136Y,2017ApJ...834..124Y}, the circumstellar environment and morphology \citep{1997ApJ...491..816R,2006AA...457.1081K,2013ApJ...770..129W,2002ApJ...581.1101H,2015ApJ...814..132L}, attempts of searching for the companion donor star \citep[][]{2004Natur.431.1069R,2009ApJ...701.1665K,2015ApJ...809..183X}, the non-detection of a Str\"{o}mgren sphere \citep{1994ApJ...431..237R,2003ApJ...590..833G,2017NatAs...1..800W}, light echo \citep[][]{2008Natur.456..617K}. However, no consensus has been reached so far as for whether the SNR originates from a DD or SD channel with competing evidences from both camps. 

%A DD scenario for Tycho has been long supported mainly by inferred uniform medium and thus unlikeliness of pre-SN activities due to some observed properties, almost spherical shape   dispute over the existence of donor star with no conclusive evidence ,  ; whereas Tycho's X-ray emission from the ejecta which Mn and Ni exists suggest SD scenario \citep{}. 

%The progenitor of Tycho has recently been considered to be surrounded by a wind-like environment \citep[][]{2016ApJ...826...34Z} and a molecular cloud \citep{2010ApJ...709.1387K,2016ApJ...823L..32W} \textbf{(\citet{2016ApJ...826...34Z} proposed a SD scenario while \citet{2017NatAs...1..800W} favored a DD scenario.).} \citet{2010ApJ...709.1387K} measured the proper motion of the forward shock at different azimuthal regions, and \citet{2016ApJ...823L..32W} found an offset of the explosion center from the geometric center. \textbf{Recent} work by \citet{2021ApJ...906L...3T} (hereafter \citetalias{2021ApJ...906L...3T}) which used archival data of proper motion observations to assess the long-term time evolution of the shock expansion, 
Tycho has recently been considered to be surrounded by a non-uniform pre-shock density azimuthally; \citet{2010ApJ...709.1387K} measured the proper motion of the forward shock at different azimuthal regions, and \citet{2016ApJ...823L..32W} found as offset of the explosion center from the geometric center. It has been suggested that the preogenitor of Tycho is further surrounded by a wind-like environment with a molecular cloud, which has led \citet{2016ApJ...826...34Z} to propose the SD scenario. However, the issue had been controversial, with a counter argument (pointing to non-association of the molecular cloud) presented by \citet{2017NatAs...1..800W} who favored the DD scenario. This was recently settled by recent work by \citet{2021ApJ...906L...3T} (hereafter \citetalias{2021ApJ...906L...3T}), who examined the long-term time evolution of the shock expansion through the proper-motion study using Chandra archival data; 
they revealed a substantial deceleration of the shock in the last couple decades, probably caused by the existence of an outer dense region of circumstellar matter (CSM), e.g., molecular clouds. Such a wind-like bubble adds another source of credibility to the SD scenario for the case of Tycho. Similar expanding bubble-shells have also been discovered in other Galactic and extragalactic SNRs \citep{2022ApJ...933..157S,2023arXiv230303456G} as well.  

In this work, we utilize the new proper motion results in an attempt to refine the picture of the environment surrounding this important historic SNR. 
First, we constrain the distance and cavity density of Tycho 
%depending on literature 
by using the forward shock (FS) and reverse shock (RS) observations simultaneously \citep{2005ApJ...634..376W,2014ApJ...780..136Y}. For a given cavity density (and the corresponding distance), we calculate the shock dynamics using hydrodynamics simulations and search for the best-fit environmental parameters, and then compare them with the evolution of the shock radius observed by \citetalias{2021ApJ...906L...3T} using a goodness-of-fit parameter ($\chi^2/\mathrm{dof}$). A few scenarios for the circumstellar environment are explored and their compatibility with the latest proper motion data are assessed. We focus primarily on the environmental aspects in this study and use fixed parameters for the SN explosion itself, which will be discussed in more details in a follow-up work. 

%We then further constrain our models in specific scenarios for the CSM as well as the prediction of the resulted spectrum and comparison with the observed spectrum. 

In Section~\ref{sec:setup}, we define models for the environment and introduce our numerical method. In Section~\ref{sec:uni}, we explore models with a uniform cavity for the environment and demonstrate the difficulty of reproducing the latest proper motion data in this kind of environments. In Section~\ref{sec:win}, we switch to the case of a wind-like cavity and search for the best-fit parameters, such as the pre-SN mass-loss rates and gas cloud densities. Section~\ref{sec:disc} is devoted to the discussion of how our best-fit models stand in the existing scenarios of Type-Ia SNR progenitors. Section~\ref{sec:sum} concludes our work. 
%In Section 2 we first explain our numerical methods which enable us to calculate SNR evolution until a few or tens of 104 yrs, and then introduce our models for the surrounding environments in this paper, i.e., models with a unifor ambient medium and those with a CSM created by the pre-SN stellar wind. In Section 3.1 and Section 3.2, we present our results from both classes of models sequentially and discuss their various implications. Section 3.3 is dedicated to the analyses of a few physical effects especially relevant to the non-thermal emission in the radiative phase, followed by a brief discussion on the future prospect of FORCE in hard X-ray studies of young and old SNRs in Section 3.4. Section 4 provides a summary of our results and concluding remarks

%\section{Observation data} \label{sec:obs}
%The explosion center is determined by tracing the recent motion. The image used is same as the data used in T+21. 

\section{Setup and models}\label{sec:setup}

\begin{figure}[ht]
    \epsscale{1.15}
    \plotone{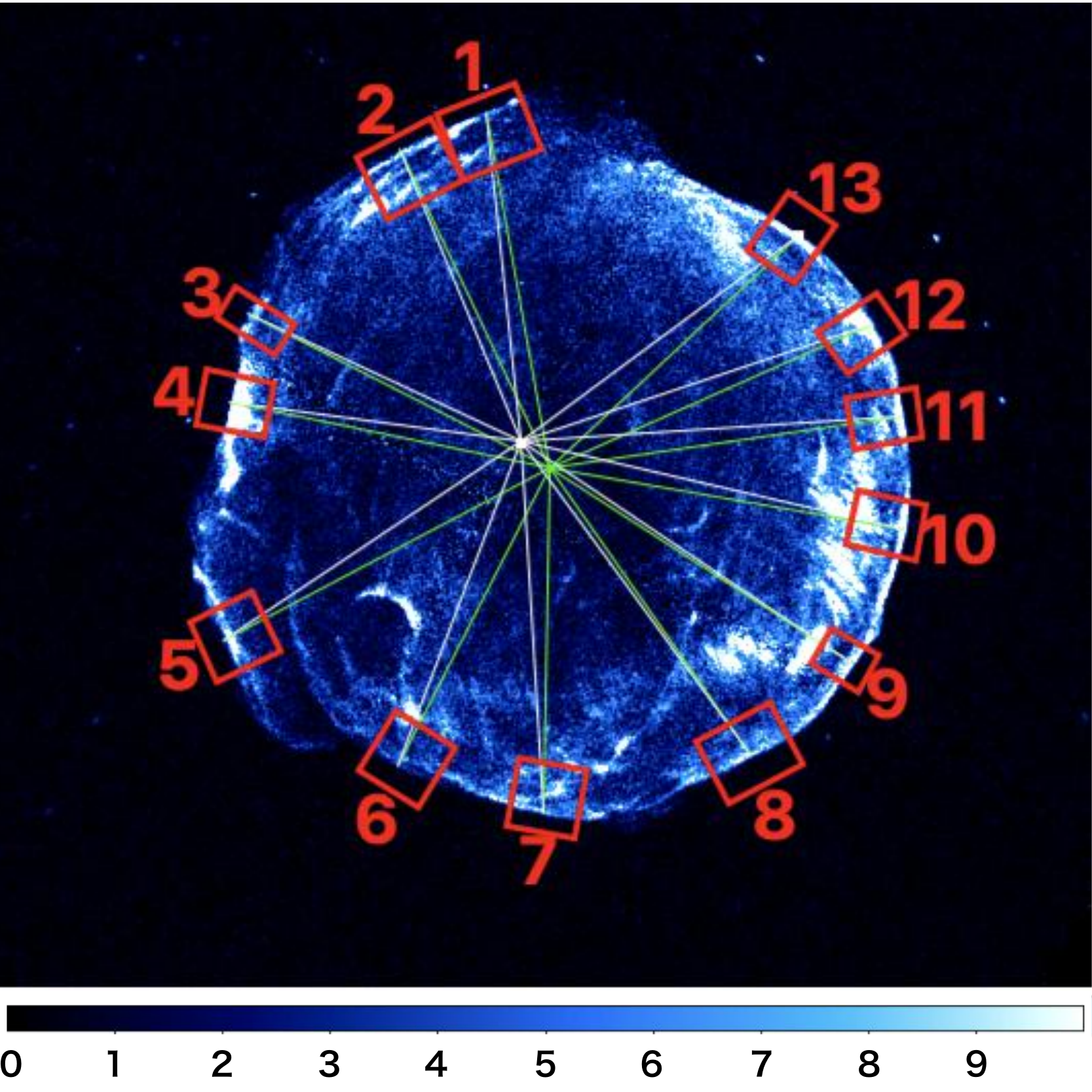}
    \caption{Chandra ACIS image of Tycho's SNR at year 2009 (reproduced from \citetalias{2021ApJ...906L...3T} data). The color bar shows the flux from each pixel in units of $10^{-9}\ \mathrm{photon\ cm^{-2}s^{-1}}$. The red rectangles are the regions we use. Green or white lines indicate the line segment connecting the forward shock in each azimuthal angle with the geometric center or pressure center, respectively.}
    \label{fig:t+21fig}
\end{figure}

\begin{figure}[ht]
    \epsscale{1.15}
    \plotone{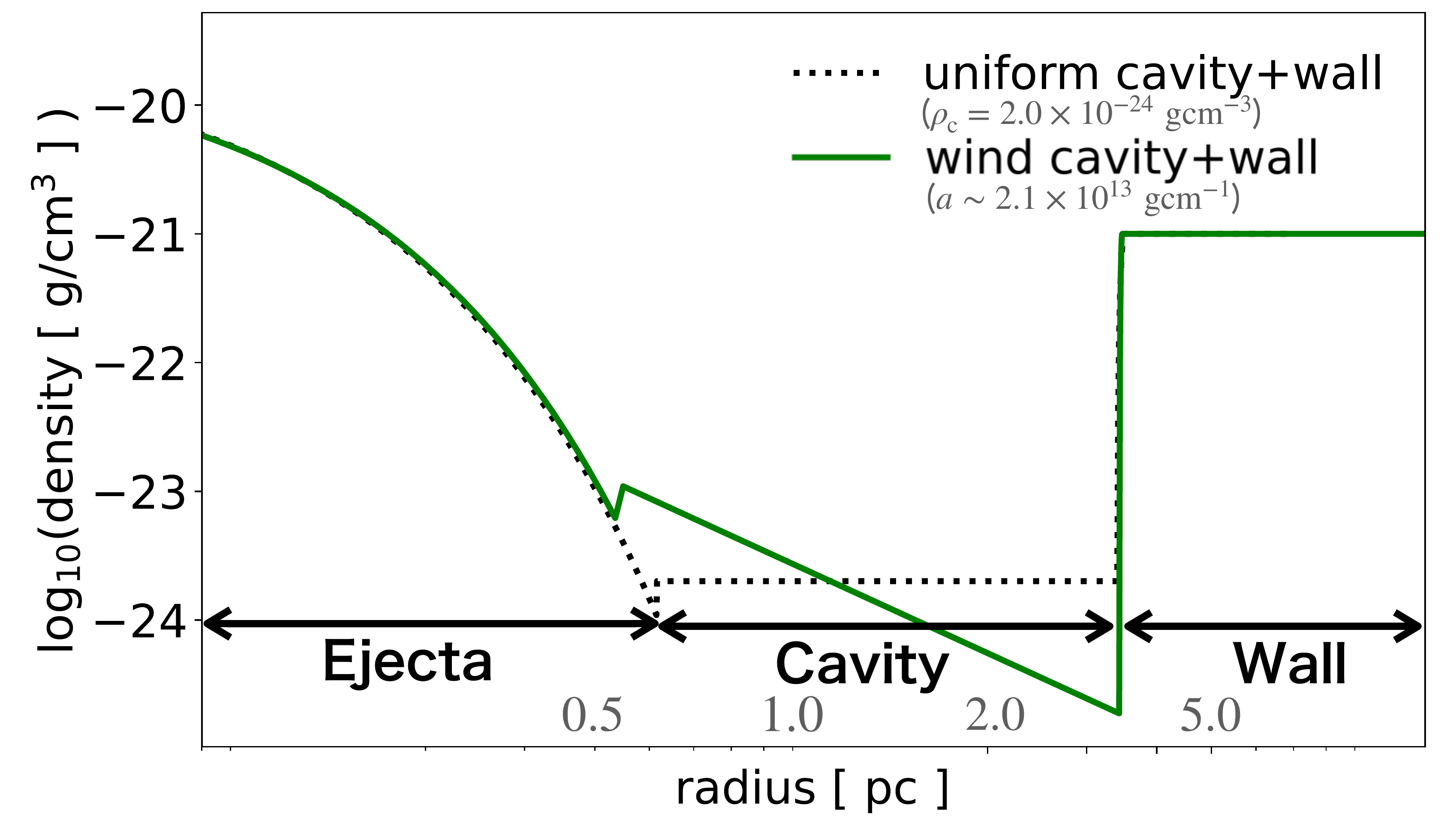}
    %\caption{Initial density profile for one of the (uniform) cavity-wall models in black lines and for one of the wind cavity-wall models in green lines, where $\rho_\mathrm{c}$=2.0$\times10^{-24}\ \mathrm{gcm^{-3}}$ and $a=10^{-}\mathrm{gcm^{-1}}$, respectively.}
    \plotone{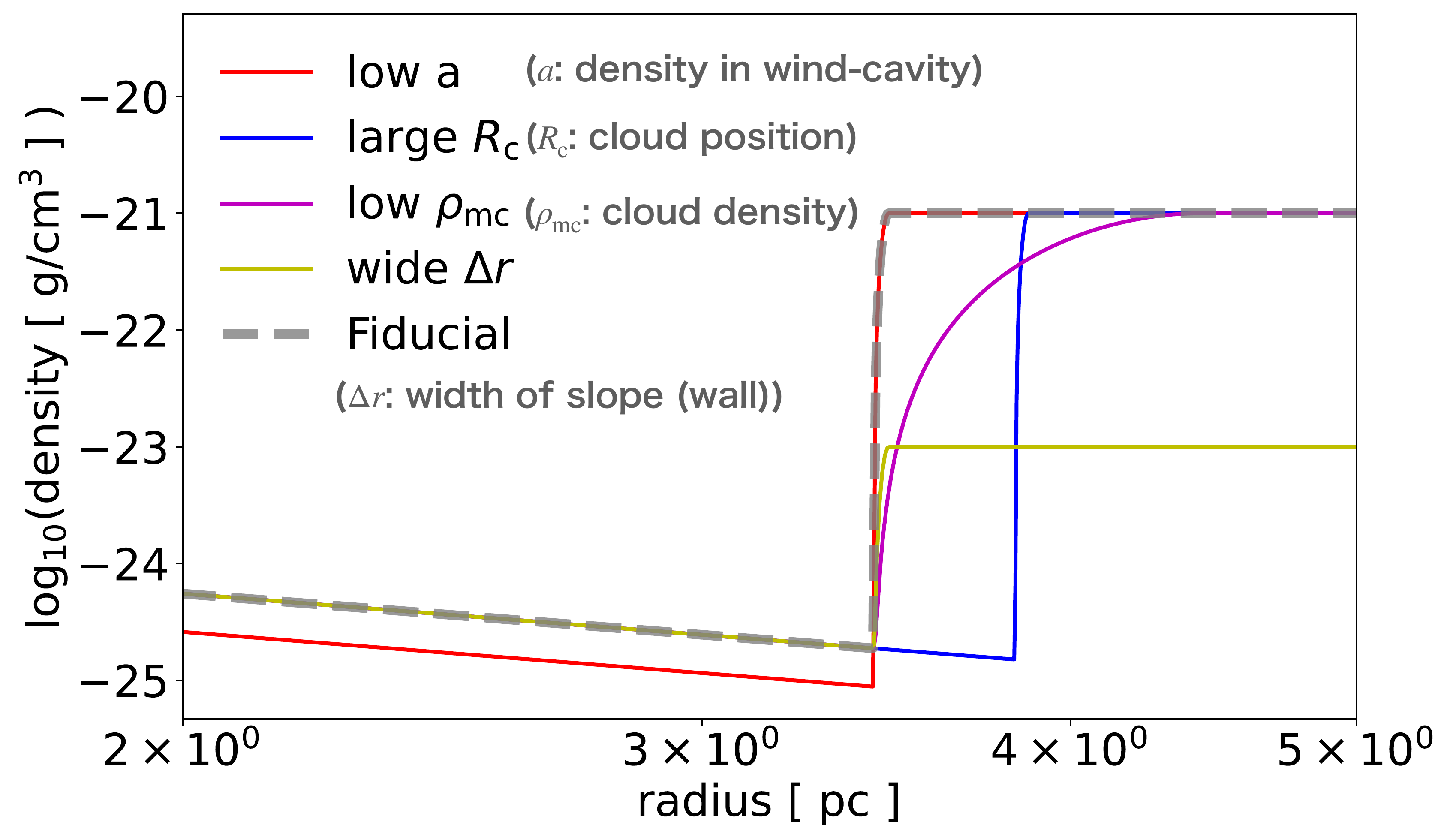}
    \caption{Upper panel: the initial density profile of the two model categories; one for the uniform-density cavity case (black line; with $\rho_\mathrm{c}$=2.0$\times10^{-24}\ \mathrm{gcm^{-3}}$), and the other for the wind-like cavity case (green line; with $a=2.1\times10^{13}\mathrm{gcm^{-1}}$, which corresponds to $\dot{M}=1.047\times10^{-4}\ \mathrm{M_\odot\ yr^{-1}}$ for $V_\mathrm{w}=250\ \mathrm{km\ s^{-1}}$). Lower panel: the initial density profiles of a wind-like cavity model, showing the effect of varying different parameters; the density in wind-cavity $a$ (red), the cloud position $R_\mathrm{c}$ (blue), cloud density $\rho_\mathrm{mc}$ (yellow), the gap between cavity and cloud $\Delta r$ (magenta). The fiducial model is shown by the grey dashed line.}
    \label{fig:d-ini}
\end{figure}

The observation data for the proper motion of the forward shock (FS)\citepalias[][in synchrotron dominated X-ray band]{2021ApJ...906L...3T} are split into four epochs in 2003, 2007, 2009 and 2015. We use the data in the year 2003 as an anchor and define positional shifts using the 2003 data as a reference point. 
%For 2003 data itself, this time 
To this end, we assume two possible locations for the explosion center, i.e., the geometric center from \citet{2005ApJ...634..376W} \citep[consistent with the one estimated from 3D proper motion of the ejecta in][]{2022ApJ...937..121M} and the one derived in \citet{2016ApJ...823L..32W} where a global pressure gradient is assumed (hereafter `pressure center'). Data in each epoch is divided into 13 azimuthal regions in a counterclockwise order from the north, i.e., Regions 1--13, as shown in Figure~\ref{fig:t+21fig} \footnote{We use the same ``position number'' as in \citet{2021ApJ...906L...3T} for the regions.}. 

%The data errorbars are calculated by averaging over the three epochs errorbars of \textit{Chandra} point spread function of shift from 2003 data as rough estimates. 

We consider two categories for the cavity-wall models (see top panel of Figure~\ref{fig:d-ini}), i.e., an inner bubble with a uniform density $\rho_0$, and one with a wind-like profile with density $\rho(r)=ar^{-2}$, each surrounded by a dense outer region. 
%cavity-wall with the density $\rho(r)=\rho_\mathrm{c}$ and wind-wall . 
The cavity-wall structure is supposed to be the result of the pre-SN progenitor activity interacting with a surrounding molecular cloud \citepalias{2021ApJ...906L...3T}. 
%, and the difference between these two models is inner sphere within clouds. 
The shock is fast when it is propagating inside the inner low-density bubble. As it starts penetrating into the dense region at a certain radius, a rapid deceleration of the shock as seen in \citetalias{2021ApJ...906L...3T} is expected. According to upper panel of Fig. 2 in \citetalias{2021ApJ...906L...3T} Regions 5--8 can be used for calibration of our models as the shock in these portions most likely have not yet started interacting with the dense cloud at year 2003 and are still inside the cavity, meaning that only one parameter is involved (i.e. the inner cavity density). 
%we can regard 2003  and  2007 data in such azimuthal angle not or least likely influenced by the dense wall, such data is explained by inner structure only, . 
Thus, in Sections~\ref{subsec:rho-d-uni},~\ref{subsec:a-d-win}, we use the 2003 and 2007 data in Region 7 and 8 (we omit Region 5 and 6, where there is no measurement for the RS radius) for constraining the Tycho's distance as a function of the cavity density. 
%and corresponding distance with models of only cavity 
After determining the distance, we move on to the models with an outer dense cloud with a density $\rho_\mathrm{mc}$, which will be discussed in Sections~\ref{subsec:r-t-uni},~\ref{subsec:r-t-win}. The bottom panel of Figure~\ref{fig:d-ini} shows the impact of each of the parameters on the density profile of our CSM model. The radial boundary of the cavity (or inner cloud radius) is $r=R_\mathrm{c}$, and the transition length between the cavity and cloud is characterized by the parameter $\Delta r$. The CSM is assumed to have, from inside to outside, a low-density cavity within $r<R_\mathrm{c}$, a density gradient connecting the cavity and cloud at a radius $R_\mathrm{c}<r<R_\mathrm{c}+\Delta r$, and a dense cloud at $r>R_\mathrm{c}+\Delta r$. For simplicity, we consider $3\times3$ patterns of ($\rho_\mathrm{mc},\Delta r$), with $\rho_\mathrm{mc}\in[10^{-23},10^{-22},10^{-21}]\ \mathrm{g\ cm^{-3}}$ and $\Delta r\in[0.04,0.4,1.0]\ \mathrm{pc}$. We also do not consider episodic mass loss which is reserved for future work in which hydrodynamic models of wind-blown bubbles will be discussed. 

For the ejecta, we use a single canonical thermonuclear explosion model with a near-Chandrasekhar mass white dwarf, i.e., an ejecta mass of 1.4~M$_\odot$, an explosion kinetic energy of 1.0$\times10^{51}\ \mathrm{erg}$, and an exponential density profile for the initial ejecta \citep{1998ApJ...497..807D}. 

%For calculating time-evolution of our models we use the \textit{CR-Hydro} code \citep{2019ApJ...876...27Y,2022ApJ...936...26K}, which calculate 1-D hydrodynamics and particle acceleration by non-linear diffuse shock acceleration simultaneously. 
The time evolution in our models is followed by a 1-D spherically symmetric hydro code \footnote{By treating each azimuthal region using independent 1-D hydro models, we are ignoring additional multi-dimensional effects such as possible clumpiness in the environment and an asymmetric thermonuclear explosion \citep[see, e.g.,][]{2022ApJ...937..121M}. The discussion of these effects is postponed to a followup work in-progress using results from fully 3-D hydro models. } \textit{VH-1} \citep[][]{2001ApJ...560..244B} on a Lagrangian grid, in which radiative cooling is accounted for using a cooling curve from \citet{1993ApJS...88..253S} \citep[see][]{2022ApJ...936...26K}. 
%This time we divide the whole (spherical) region into 13 segmented regions and assume that we can regard each region homogeneous. Multi-dimensional effect will provide us complicated structure of circumstellar medium or asymmetry of explosion energy \citep[e.g.,][]{2022ApJ...937..121M}. 

\section{Uniform cavity models}\label{sec:uni}
\subsection{Calibration of cavity parameters and estimates of the distance}\label{subsec:rho-d-uni}
\begin{figure*}[ht]
    \epsscale{1.15}
    %\plotone{shock2003-D_2_W+05-re7-1.pdf}
    \gridline{\fig{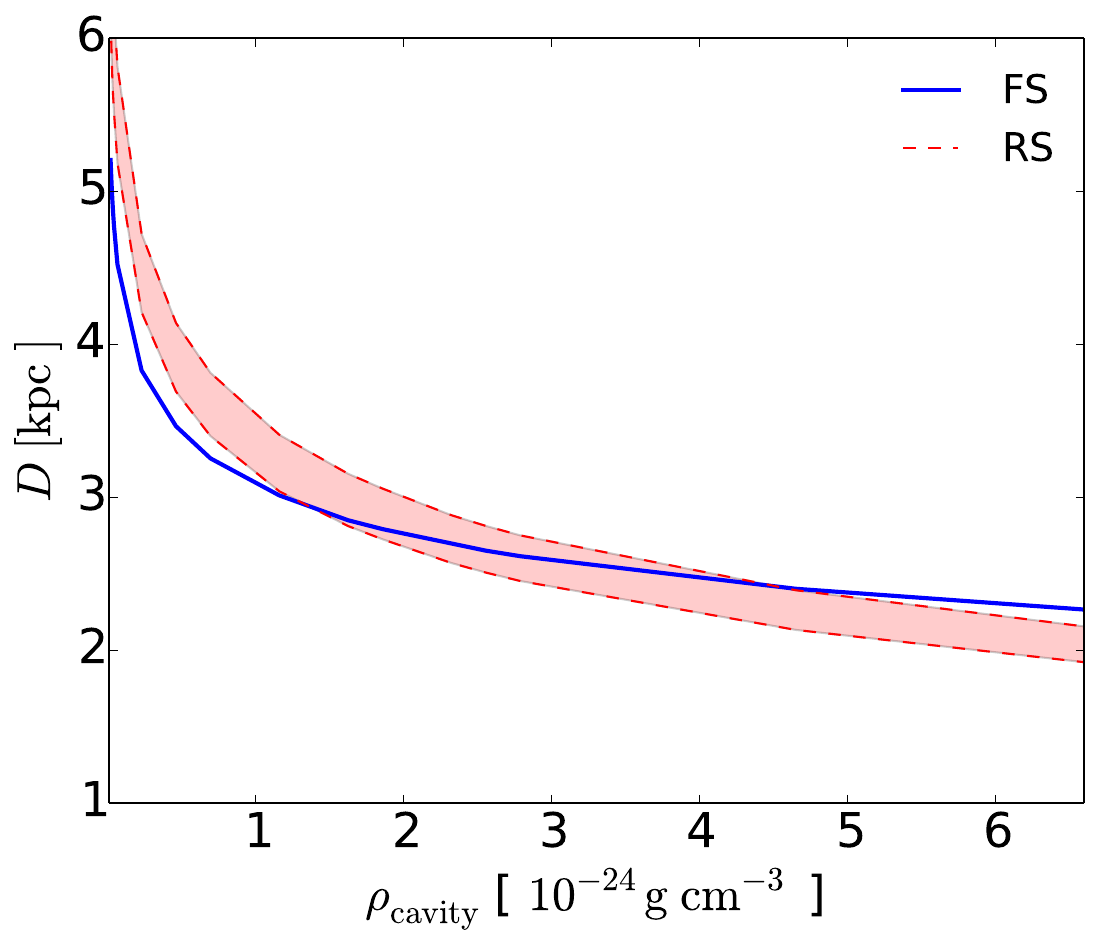}{0.45\textwidth}{(a) Region 7 (Geometric center)}
    \fig{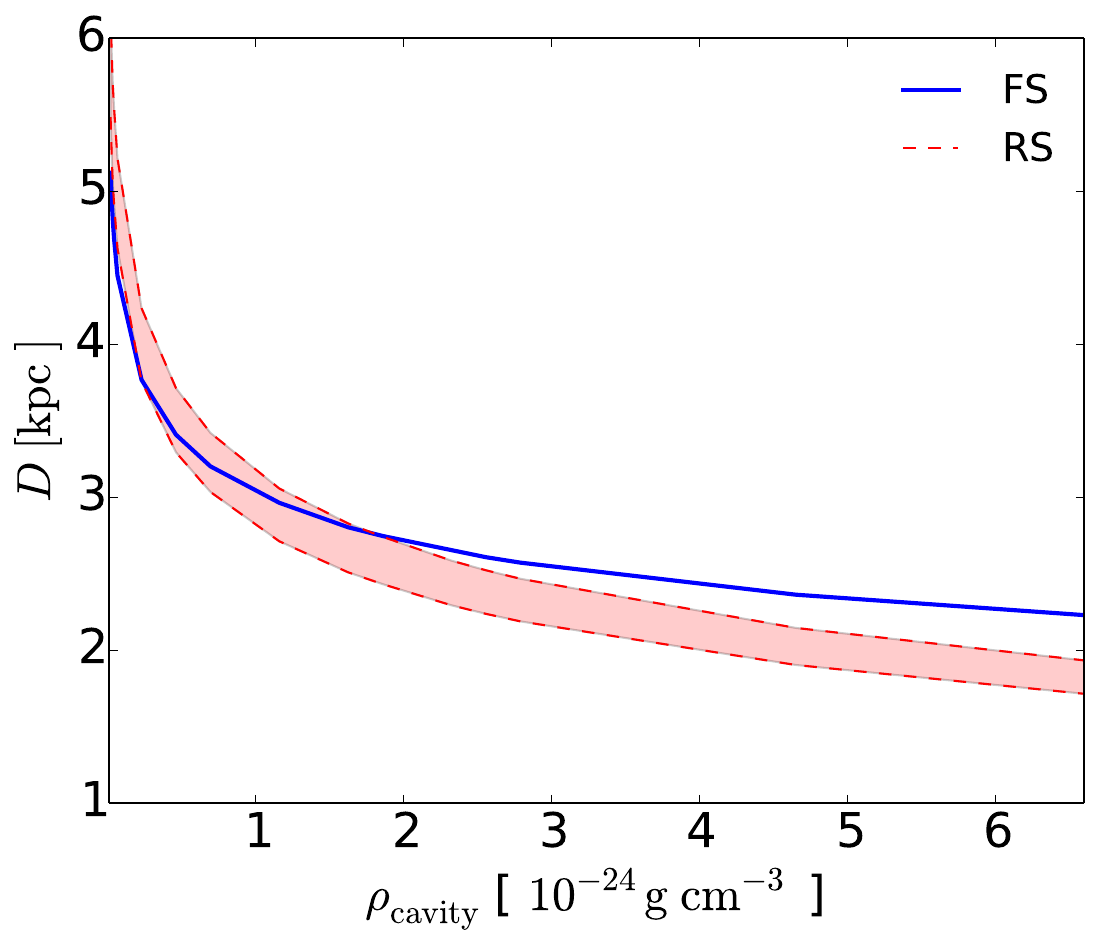}{0.45\textwidth}{(b) Region 8 (Geometric center)}}
    \gridline{\fig{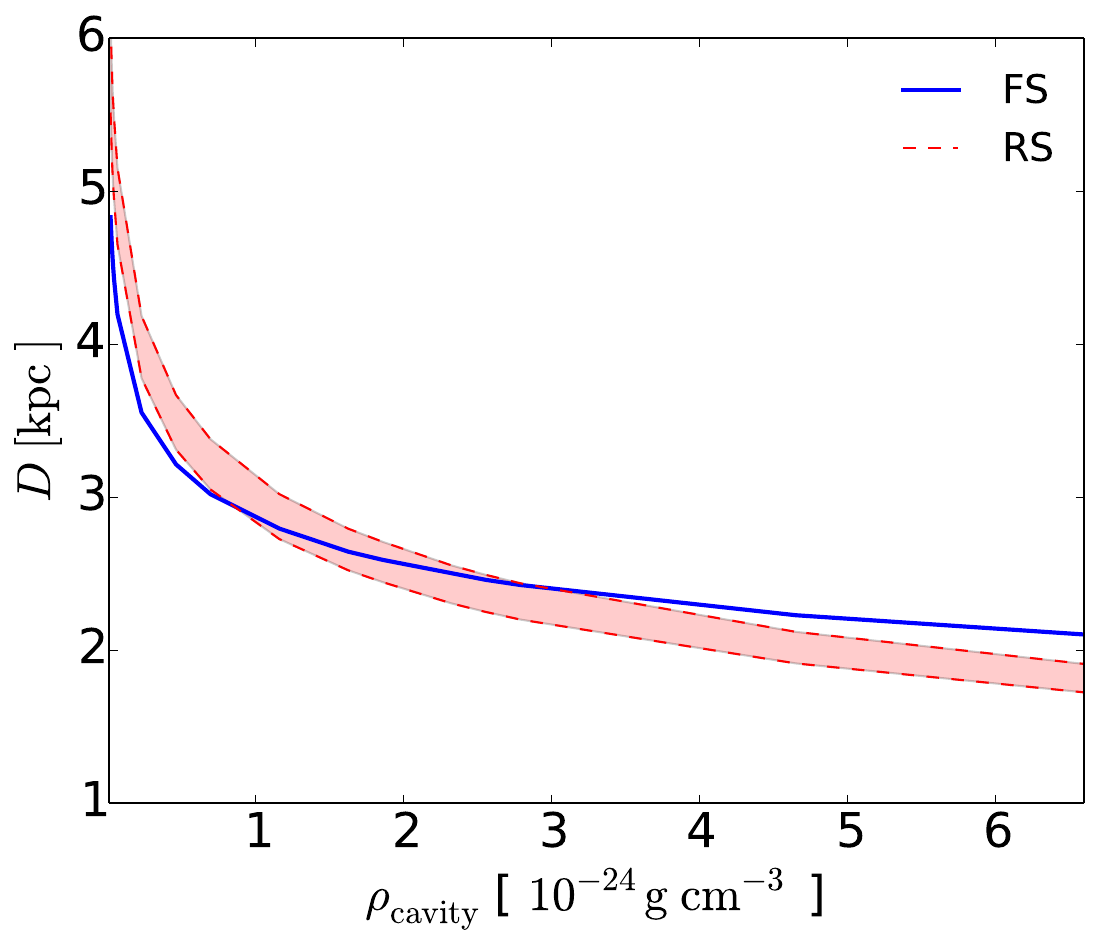}{0.45\textwidth}{(c) Region 7 (Pressure center)}
    \fig{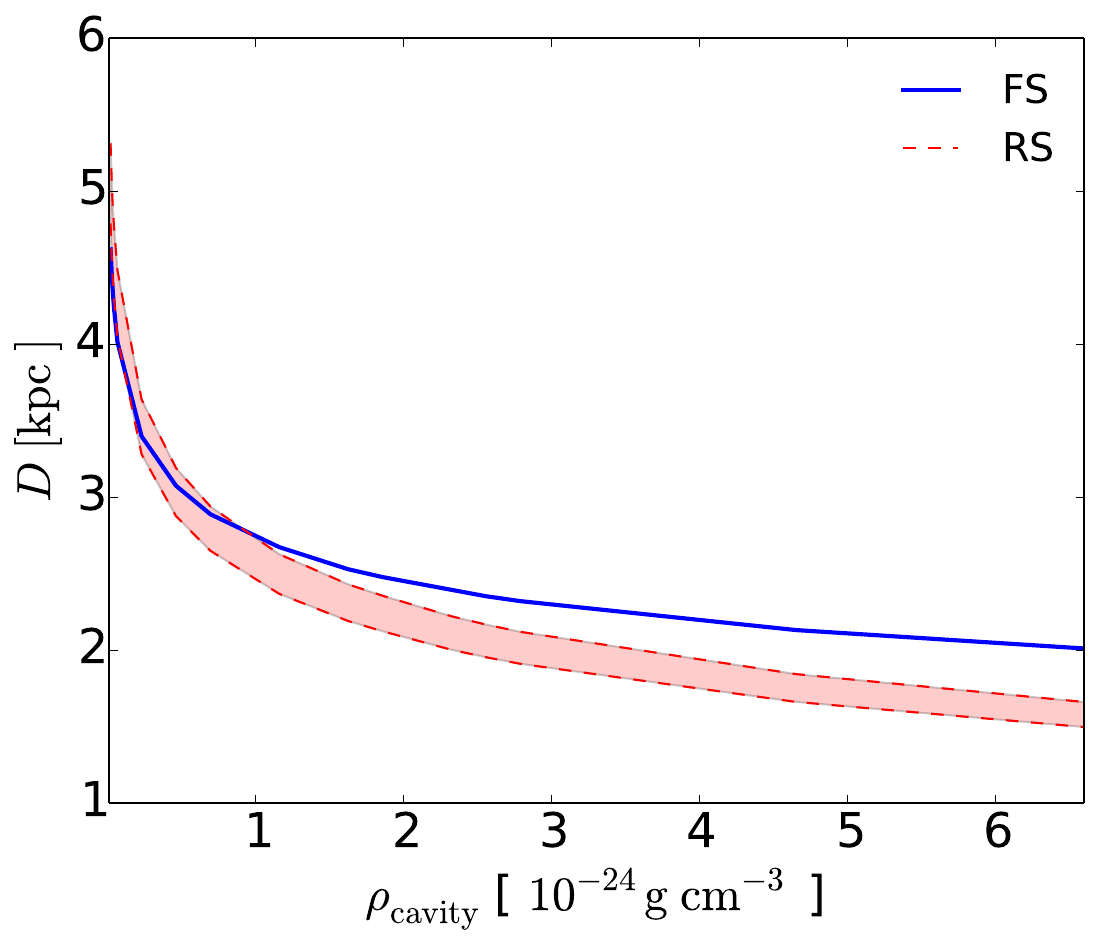}{0.45\textwidth}{(d) Region 8 (Pressure center)}}
    \caption{Cavity density versus distance results from our hydro models constrained by the FS (blue lines) and RS (red regions) observation data. The ranges where the blue lines overlap with the red shaded area are consistent with both the FS and RS data simultaneously. The left and right panels correspond to the results from Region 7 and Region 8, where the geometric (pressure gradient) explosion center is assumed in the upper (bottom) panels.
    }
    \label{fig:n-D}
\end{figure*}

We first consider a uniform ambient environment with a constant density $\rho_\mathrm{c}$ ignoring the effect of an outer dense region, and focus on the relation between $\rho_\mathrm{c}$ and the distance $D$ of Tycho from Earth. 
Despite the many works done so far \citep[see][Fig.6]{2010ApJ...725..894H}, there still remain large uncertainties among the literature values on the distance to Tycho. A precise knowledge of the distance is especially crucial to break degeneracy during comparison between the observation data (in angular scale) and hydro models (in physical scale). From the observed angular radius $\theta$ of the FS using the synchrotron-dominated X-ray emission \citepalias[][]{2021ApJ...906L...3T}, we can calculate the distance as $R_\mathrm{sim}(\rho_\mathrm{c})/\theta$ from the model radius $R_\mathrm{sim}(\rho_\mathrm{c})$ assuming a certain value for the density of the ambient medium. Furthermore, we can further narrow down the value by performing a similar calculation 
%a similar quantity $R_\mathrm{sim}(\rho_\mathrm{c})/\theta$ 
but for the reverse shock (RS).

Likewise, there exists an uncertainty on the density estimations \citepalias[see][and references therein]{2021ApJ...906L...3T}. 
For the angular radii $\theta$ of the FS and RS at each region, we use the observation results from \citet{2005ApJ...634..376W} as a reference. For simplicity, we use the K$\alpha$-peak as in \citet{2005ApJ...634..376W} for pinning down the RS position, although we are aware that there is evidence for some regions that the RS radius can be smaller if the Fe-K$\beta$-peak is used as a beacon for the RS location instead of K$\alpha$ by considering the post-shock ionization length scale \citep[e.g.,][]{2014ApJ...780..136Y}. 
%\begin{bfseries}
The analysis of the Fe-K$\beta$ peak, however, has been performed only for the NW region, and such analyses for other azimuthal regions are not yet available. %The data of Fe-K$\beta$ peak in \citet{2014ApJ...780..136Y} is shown only for NW region, and it is hard to convert into other azimuthal regions without introducing further model-dependent uncertainties. %convert into other values of densities because ionization of heavy ions will be needed to simulate.
%\end{bfseries} 
We anticipate that an extension to the usage of Fe-K$\beta$ data can be done consistently when an observation along the full azimuthal circle will be performed in the future. 
%\footnote{We note that by K$\alpha$-K$\beta$ gap \citep{2014ApJ...780..136Y} the data of RS might be smaller in some azimuthal regions.}
%especially in a low-density wind. } %\footnote{We suppose K$\alpha$--K$\beta$ width constant over different $a$ assuming that ionization degree is only weakly dependent on wind properties.}. 

In Figure~\ref{fig:n-D}, we plot the density-distance relations by using the FS and RS measurements in the year 2003 as constrained for Region 7 and 8 (in which the SNR has presumably started interacting with the dense region well after 2003; see Section~\ref{sec:setup}). The overlapping regimes in the $\rho_\mathrm{c}-D$ parameter space, i.e., where the blue lines overlap with the red shaded area, show the allowed ranges, being consistent with both the FS and RS data simultaneously. 
%especially in Region 7 and 8 as representative regions least likely. 
We hence obtain the estimates of $D$ and $\rho_\mathrm{c}$ as $\sim3$~kpc and $\sim2\times10^{-24}\ \mathrm{g\ cm^{-3}}$, respectively. The results are shown under two assumptions on the explosion center location as well, as discussed above. 
%These results from the FS data are robust in each way of center; whereas in Section~\ref{subsec:r-t-uni} we do not apparently use RS constraints since determination of RS position is difficult. 

\subsection{Proper motion simulation}\label{subsec:r-t-uni}

\begin{figure}[ht]
    \epsscale{1.15}
    \plotone{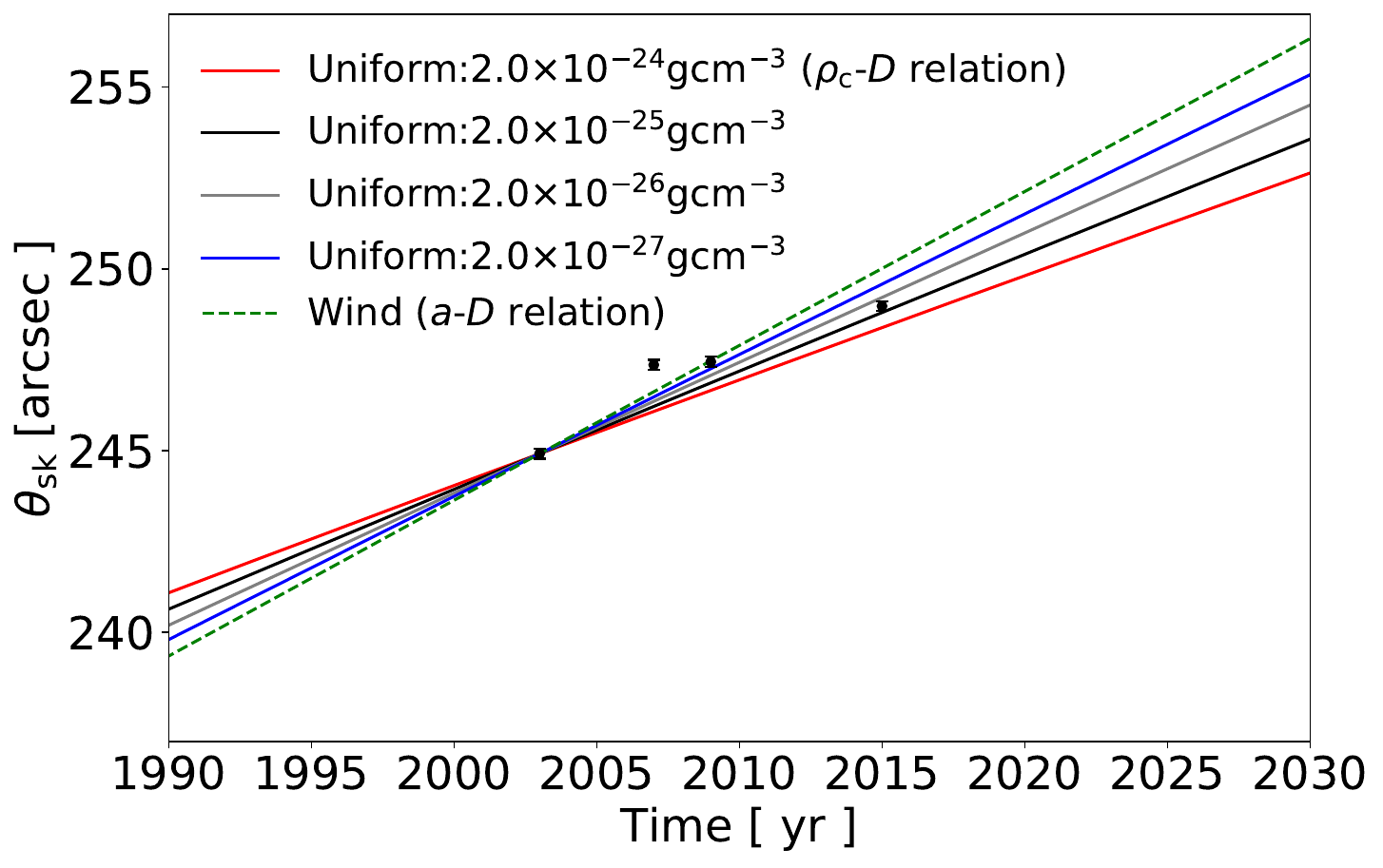}
    %\fig{Re8-2_all_rsk3_tall_uniwi.pdf}{0.45\textwidth}{(b) Region 8(Geometric center)}}
    %\gridline{\fig{Re7-1_all_rsk3_tall_uniwi.pdf}{0.45\textwidth}{(c) Region 7(Pressure center)}
    %\fig{Re8-1_all_rsk3_tall_uniwi.pdf}{0.45\textwidth}{(d) Region 8(Pressure center)}}
    \caption{Time-evolution of the FS radius for the case of a uniform cavity compared with data in Region 7, with the geometric explosion center assumed. The red line shows a model with parameters consistent with both the FS and RS radii at 2003. The others assume smaller cavity densities with corresponding distances. A wind-like model is also shown with the green dotted line. The choice of explosion center is not critical to the result, and a similar result is obtained for Region 8.}
    \label{fig:n-shift-1}
\end{figure}

Using the relation obtained in Section~\ref{subsec:rho-d-uni}, here we perform hydrodynamic simulations with different environment models and compare them with the data for the FS position to assess their compatibility with the observations. 

Figure~\ref{fig:n-shift-1} shows the model evolution of the FS radius for an environment with a constant density $\rho_\mathrm{c}$. Here, without the complexity from cloud interaction, we can first compare a model consistent with the $\rho_\mathrm{c}$--$D$ relation obtained above (red line) with the 2003--2015 data. The model clearly underpredicts the angular size of Tycho from 2007 to 2015 data. An ad hoc attempt to realize a faster expansion by using a smaller ambient density (black, grey and blue lines) fails to account for the discrepancy, since the larger shock velocity in a more tenuous medium is offset by a larger required distance as shown in Figure~\ref{fig:n-D}, so that the apparent angular velocity (or size) is only moderately affected as we can see in Figure~\ref{fig:n-shift-1}. 
%which is divided by distance $D$ (i.e., the slope in Figure~\ref{fig:n-D}) due to the declining trend of $D$ with smaller density .    
Instead, when a wind-like cavity is incorporated (dashed green line, see Section \ref{sec:win} for details), 
%case after similar steps.  In uniform cavity case, as the smaller density is assumed, In other hands, 
the $\rho \propto r^{-2}$ density structure in the ambient environment helps realize an overall faster angular expansion to reconcile better with the observed size at each epoch, giving support to a wind-like environment around Tycho in its earlier evolution stage before its blastwave hits the dense wall. This result from fitting to data in Regions 7 and 8 provides a motivation to explore wind-like models for the environment of Tycho, and study their implications on its progenitor system in the next Sections. 
%even smaller density around the timing of the data ($\sim$2003 yr), leading to more reasonable match with the observation. 
%This informs us the fact that Region 7 and 8 data support rather wind-cavity case which we will see in the next section. 

\section{Wind cavity models}\label{sec:win}
The existence of a progenitor wind for a Type Ia SNR is supported by some previous studies (see Section~\ref{sec:intro}). The failure of the models with a uniform cavity on explaining the time evolution of the proper motion of Tycho as we have seen above leads us to explore models invoking a pre-SN wind-blown bubble. We will separate the discussion into an isotropic wind and anisotropic wind cases. 

\subsection{Calibration of cavity parameters and distance estimations}\label{subsec:a-d-win}
\begin{figure*}[ht]
    \epsscale{1.15}
   %\gridline{\fig{shock2003-D_4re7_1_2.pdf}{0.25\textwidth}{(a) Region 7 constrain}
    %\fig{shock2003-D_4re8_1_2.pdf}{0.25\textwidth}{(b) Region 8 constrain}}
    %\gridline{\fig{shock2003-D_4re78_3_2.pdf}{0.25\textwidth}{(c) Constraints for $D$ }
    %\fig{shock2003-D_4re78_2_2.pdf}{0.25\textwidth}{(d) Rough estimates for $a$}}
    \gridline{\fig{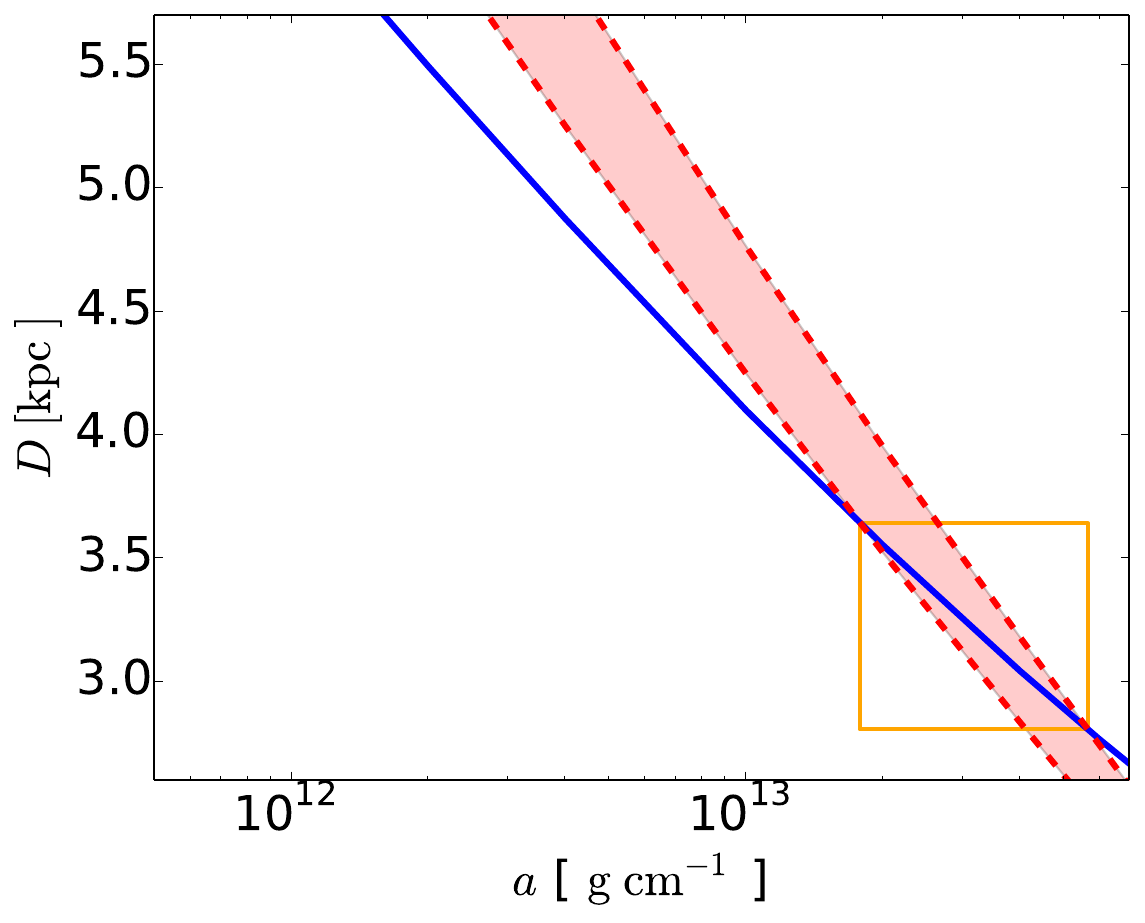}{0.45\textwidth}{(a) Region 7}
    \fig{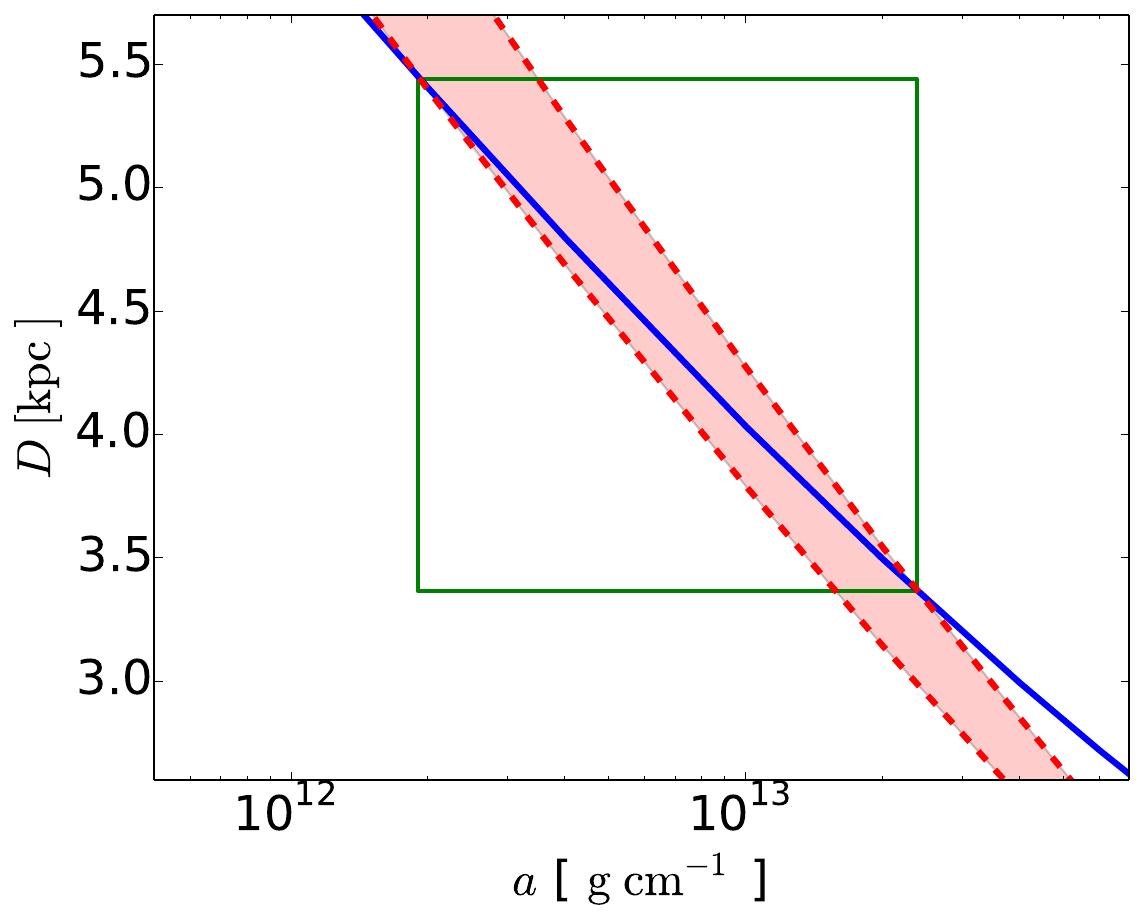}{0.45\textwidth}{(b) Region 8}}
    \gridline{\fig{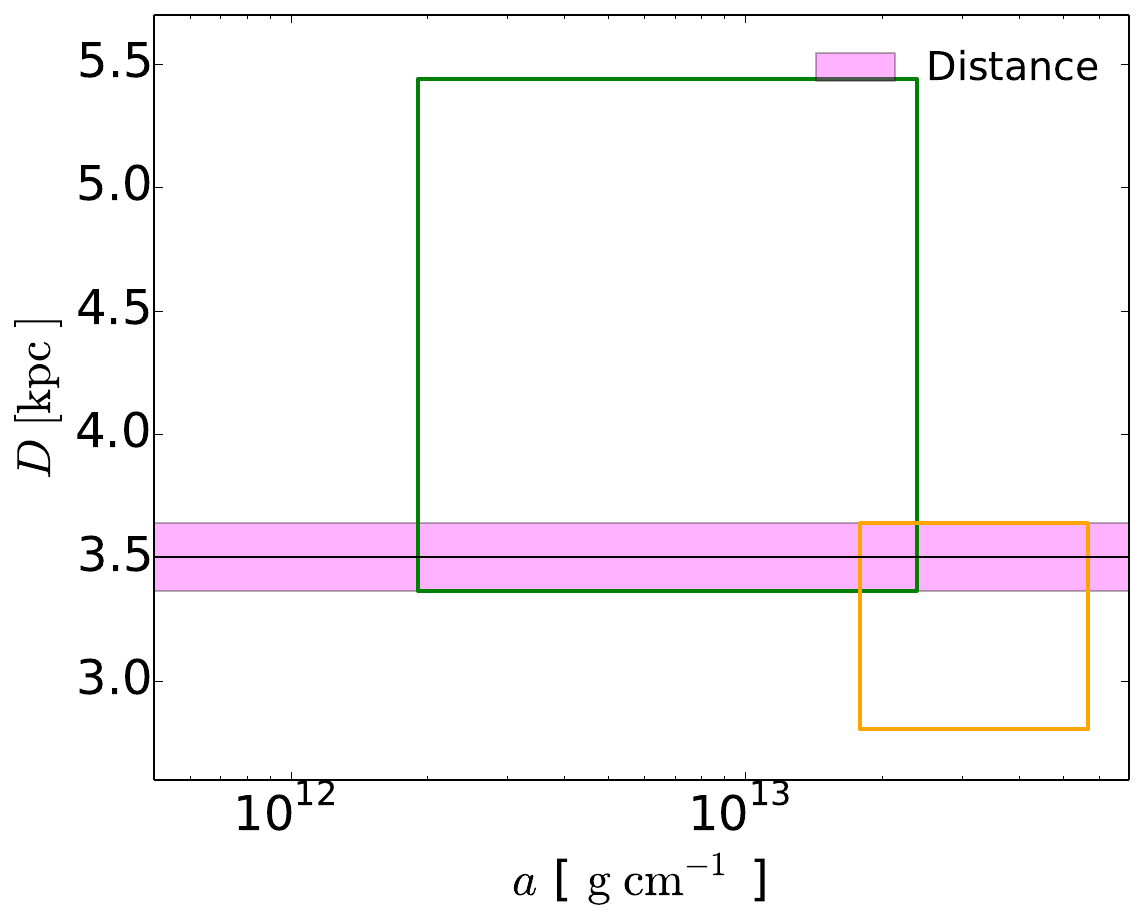}{0.45\textwidth}{(c) Combined constraints on $D$}
    \fig{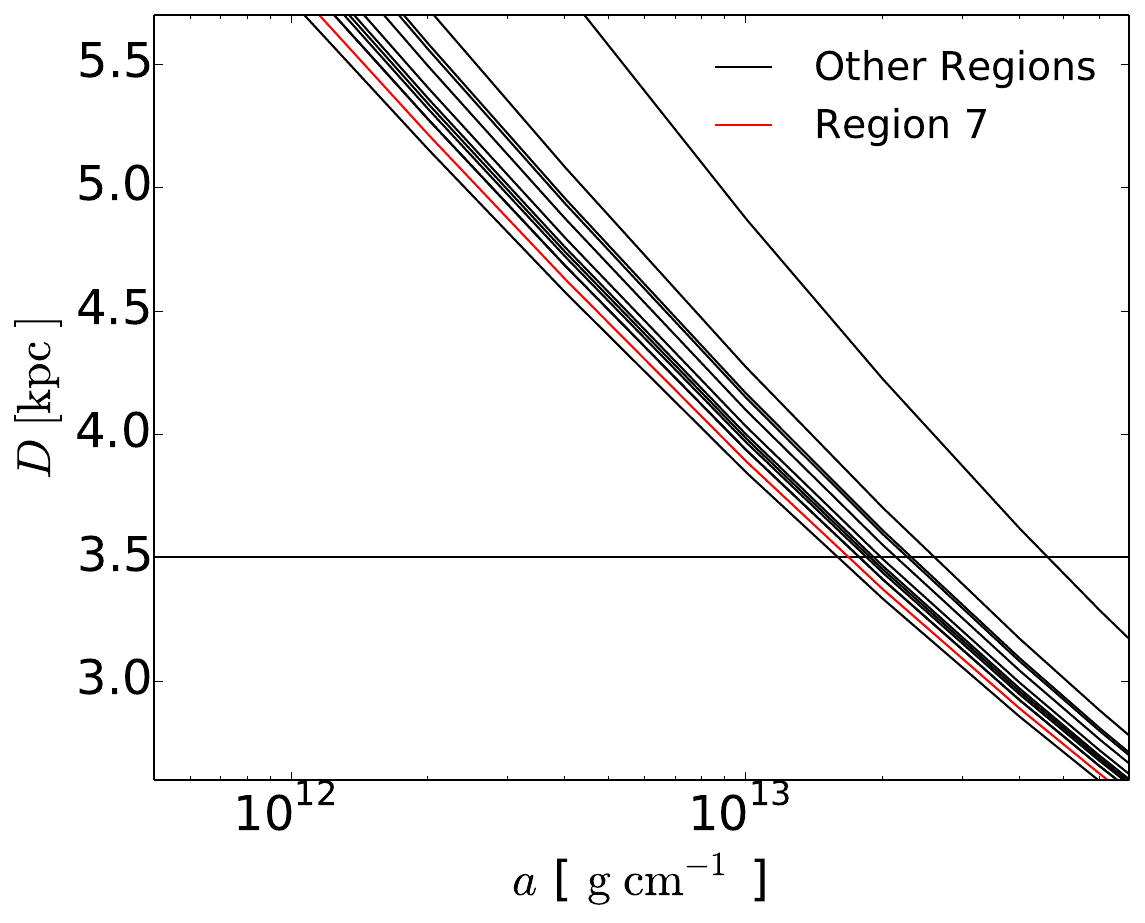}{0.45\textwidth}{(d) Rough estimates of $a$ in all regions}}
    %\plotone{shock2003-D_4re7_1.pdf}
    \caption{Panel (a) and (b): Wind parameter $a$ versus distance from our hydro models constrained by the FS (blue lines) and RS (red regions) data for Region 7  (a) and Region 8 (b). The orange and green boxes indicate the parameter range where a model is consistent with both the FS and RS data in Region 7 and 8 respectively, with the overall constraint on $a$ and $D$ shown in panel (c). %(c): Constraints in $a$-$D$ parameter space (magenta bar) from (a),(b). 
    Panel (d): Rough estimates for $a$ in each region obtained by comparing the hydro models with the FS radii at the 2003 epoch in each region, but without any cloud interaction accounted for. The horizontal line is the median of the best-fit distance $D$ from panel (c). The geometric explosion center is assumed for all results shown here.}% (e-h): same as in Panel (a-d), but using pressure center instead of geometric center when meauring data.}
    \label{fig:a-D-1}
\end{figure*}
%\begin{figure}[ht]
   % \epsscale{1.15}
   %\gridline{\fig{shock2003-D_4re7_1.pdf}{0.25\textwidth}{(a) Region 7 constrain}
    %\fig{shock2003-D_4re8_1.pdf}{0.25\textwidth}{(b) Region 8 constrain}}
    %\gridline{\fig{shock2003-D_4re78_3.pdf}{0.25\textwidth}{(c) Constraints for $D$}
    %\fig{shock2003-D_4re78_2.pdf}{0.25\textwidth}{(d) Rough estimates for $a$}}
    %\plotone{shock2003-D_4re7_1.pdf}
    %\caption{The same figure as in Figure~\ref{fig:a-D-1}, but using pressure center instead of geometric center when meauring data. }
    %\label{fig:a-D-2}
%\end{figure}

For the wind-cavity models, we follow similar steps as above but now with a wind parameter $a\equiv\dot{M}/(4\pi V_\mathrm{w})$ instead of $\rho_\mathrm{c}$, where $\dot{M}$ is the mass-loss rate and scale by the reference wind velocity ($V_\mathrm{w}$) taken as 250 $\mathrm{km\ s^{-1}}$ according to \citet{2016ApJ...826...34Z}. From the FS and RS data at Regions 7 and  8 again, we obtain results on the $a$ versus $D$ plane (Figure~\ref{fig:a-D-1}(a,b)) and thus constrain $D$. A unique range of $D$ is determined from the simultaneous fit to both Regions (see the magenta band in Figure~\ref{fig:a-D-1}(c)). This estimate of $D$ ranges over a narrow window of $3.37-3.64$ kpc ($3.52-3.78$ kpc if the pressure center is assumed). Hereafter, we will use the median of this range as our estimate of $D$ for the analysis in all regions, which is reasonable given the small allowed range.
%is consistent with distance estimate ($\sim$ 2.5 kpc) of the CO associated clouds measured from $v_\mathrm{LSR}$ \citep{2016ApJ...826...34Z}. 

Using the obtained estimate of $D$, the appropriate wind parameter $a=a_\mathrm{region}$ in other regions can be estimated (Figure~\ref{fig:a-D-1}(d)) using the relation $R_\mathrm{sim,FS}(a_\mathrm{region})=D\theta_\mathrm{region}$. These values are obtained under the assumption that the FS has not hit the dense cloud before the 2003 epoch in all regions, which is not guaranteed to be true.
%as well these regions also explained by a single parameter which means that cloud interaction is after 2003 data and so we can determine wind density $a$ by adapting the same distance for these regions. 
As a result, $a_\mathrm{region}$ ($a\sim$a few $10^{13}\mathrm{g\ cm^{-1}}$; $\dot{M}\sim10^{-4}\ \mathrm{M_\odot/yr}$ and $V_w=250\ \mathrm{km/s}$) values serve only as rough order estimates. We can see that the estimated wind parameter $a_\mathrm{region}$ is around 100 times larger than that predicted for a normal accretion wind from the well-known Hachisu model (see Section~\ref{subsec:disc-wind-explain} for the discussion on order estimate of the wind). 

\subsection{Proper motion simulation}\label{subsec:r-t-win}

Likewise to Section~\ref{subsec:r-t-uni}, we perform hydrodynamic simulations under various environment parameters to compare with the proper motion data using the distance $D$ obtained in Section~\ref{subsec:a-d-win}. 
This time, we will investigate an isotropic wind (Section~\ref{subsubsec:iso}) and anisotropic wind case (Section~\ref{subsubsec:aniso}) in the following.

To quantitatively compare the model angular radius with the data, we use $\chi^2$ for the goodness-of-fit, i.e., %(y_i-m_i)^2
\begin{equation}
    \chi^2=\sum_{i=1}^4 \frac{(y_i-m_i)^2}{\sigma^2},
\end{equation}
where $i=1,2,3,4$ is the data number corresponding to the 2003, 2007, 2009 and 2015 epochs respectively, $y_i$ are the observed values,
$m_i$ are the values from our model, and $\sigma$ is the rough ``average error'' to make $\chi^2$ a reasonable value by $\sigma=(\alpha_2+\alpha_3+\alpha_4)/3$ where $\alpha_i$ is the error accompanied with the data of positional shifts $y_i-y_1$.%where we use the same value over $i=1-4$ since we do not know the true values. }%where we use $\sigma_i=(\delta_1+\delta_2+\delta_3)/3$ with the error $\delta_i$ accompanied with positional shifts $y_i-y_1$.} 

\subsubsection{Case with an isotropic wind}\label{subsubsec:iso}

\begin{figure*}[ht]
    \epsscale{1.15}
    \gridline{\fig{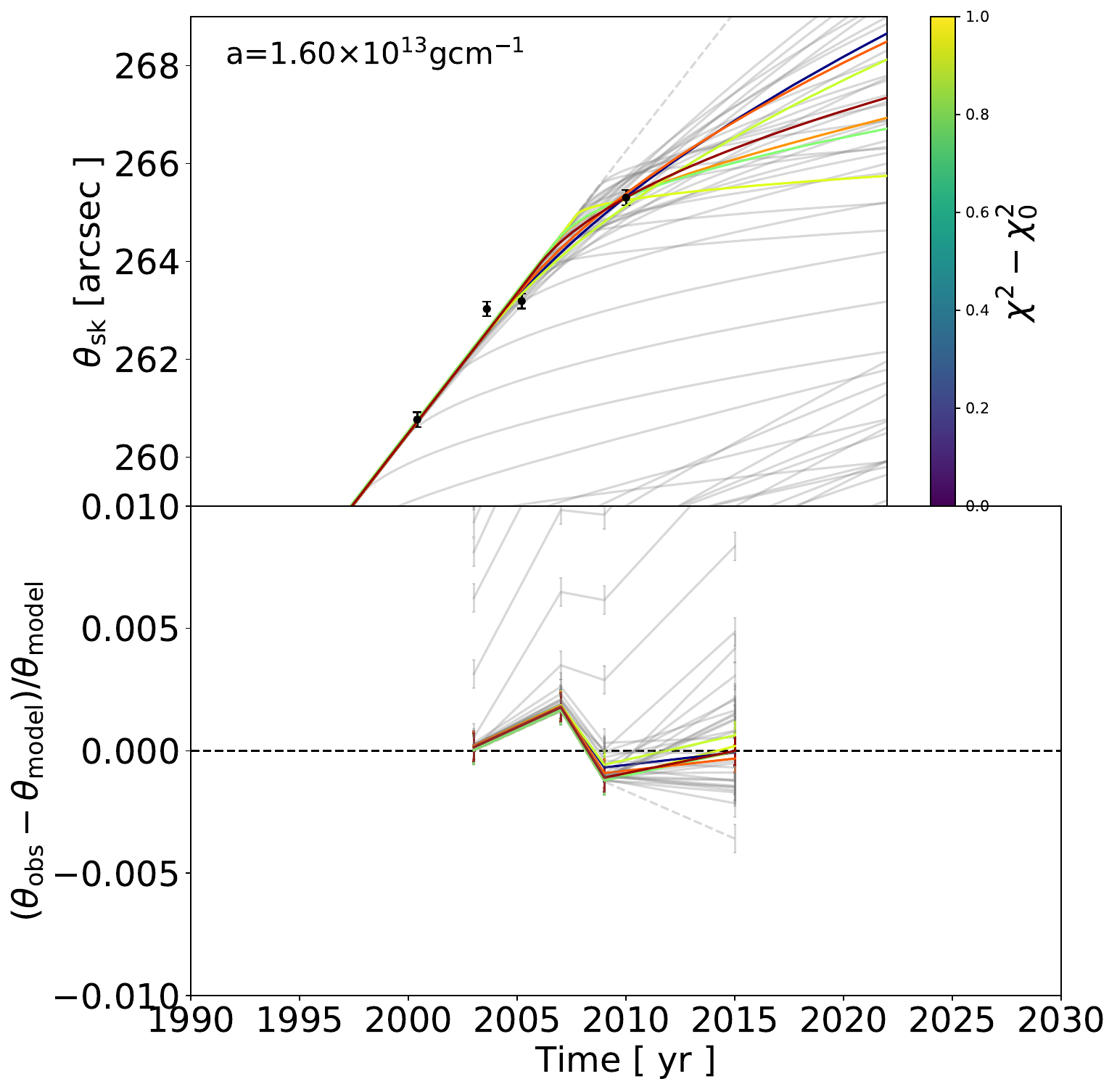}{0.5\textwidth}{(a) Region 5}
    \fig{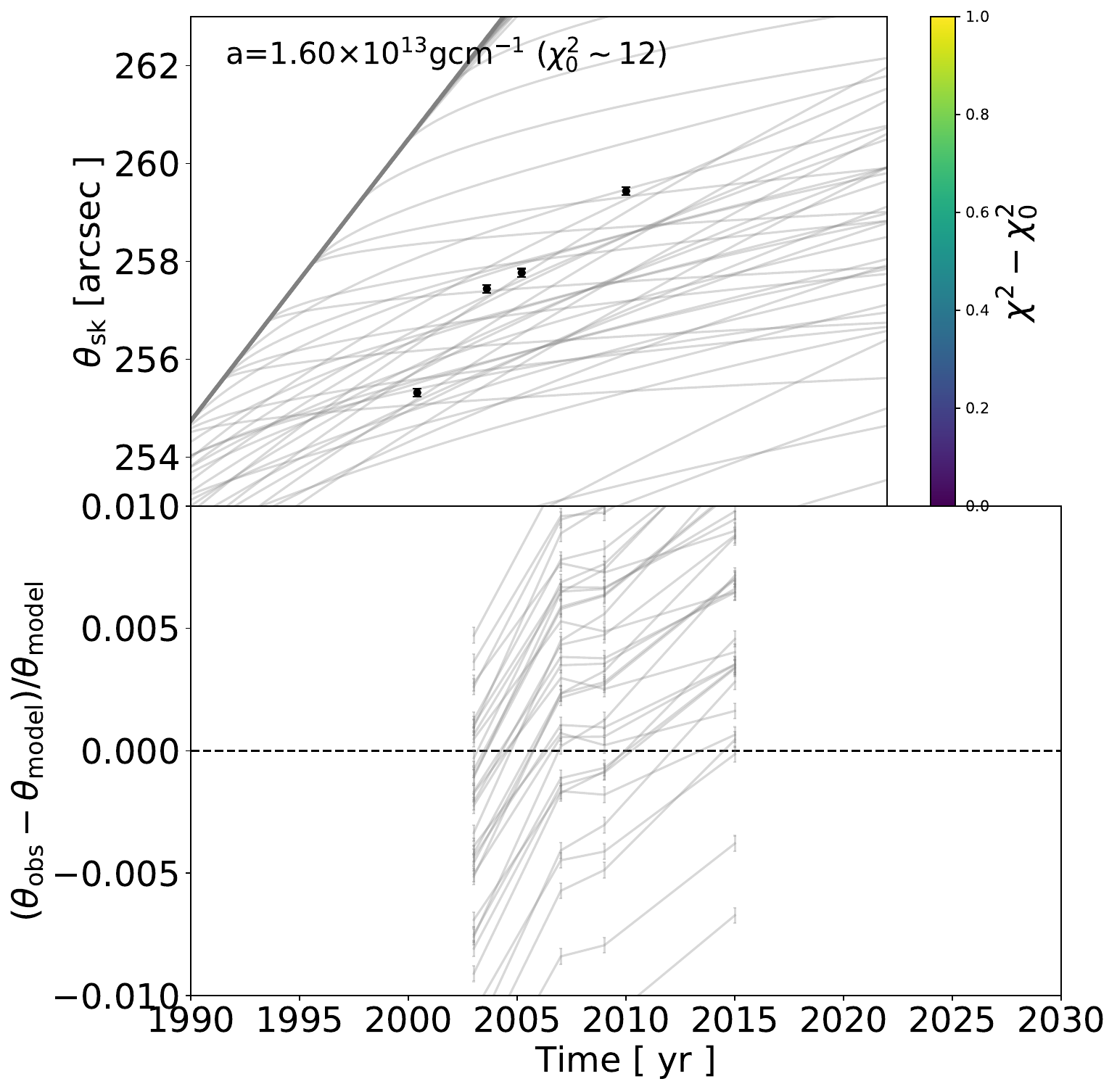}{0.5\textwidth}{(b) Region 9}}
    \caption{Time-evolution of the FS radius assuming an isotropic wind-like environment in Region 5 (Panel (a)) and 9 (Panel (b)), respectively. The lines are colored according to the color map if $\chi^2\lesssim\chi_0^2+1$, or in grey otherwise, and $\chi_0^2$ is the minimum $\chi^2$ obtained. Note that here we use the $\chi_0^2$ value obtained for Region 5 for the colorbars in both panels, i.e., $\sim$12.2 in both panels ($\chi_0^2\sim175$ in Region 9) for comparison purpose. The geometric explosion center is assumed for all results shown here.}
    \label{fig:a-shift-iso}
\end{figure*}

In this set of models, we fix the wind parameter $a$ to a unique and common value for all azimuthal angles. 
The FS angular radii over the azimuthal circle measured in the 2003 epoch are found to vary from $\sim200$~arcsec to $\sim260$~arcsec, and the maximum radius is recorded at Region 5. 
%We pick the wind parameter $a=a_\mathrm{region}$ is inversely correlated to data $\theta_\mathrm{region}$ over different regions because we compare the data $\theta_\mathrm{region}$ versus $R_\mathrm{sim}(a_\mathrm{region})/D$.
% in this case is $a\gtrsim a_\mathrm{region,5}$ where $a_\mathrm{region,5}$ is $a_\mathrm{region}$ using the value of Region 5 (Figure~\ref{fig:a-shift-iso}(a)). 
As a test, we pick the wind parameter obtained for Region 5, i.e., $a=a_\mathrm{region,5}$, and apply it to the other regions to see if satisfactory fits can be obtained with a variable cloud parameter. 
%obtain well-fitted models for Region 5 (Figure~\ref{fig:a-shift-iso}(a)). In other hand, 
The fitting in the other regions turns out to be difficult using the wind model best-fitted for Region 5, mainly because the slope between the 2003--2007 epoch is so steep that even models with the FS interacting with the dense cloud from well before 2003 cannot explain the data. An example for Region 9 is shown in Figure~\ref{fig:a-shift-iso}(b). Obviously, a single set of wind parameter (isotropic wind) for Tycho's surrounding is falling short of being able to explain the overall dynamics of the shock around the azimuth. Therefore, we next propose another set of models with an anisotropic wind, which use a variable wind parameter as a function of $\theta$, to see if we can see an improvement to the fits. 

\subsubsection{Case with an anisotropic wind}\label{subsubsec:aniso}
\begin{figure}[ht]
    \epsscale{1.25}
    \plotone{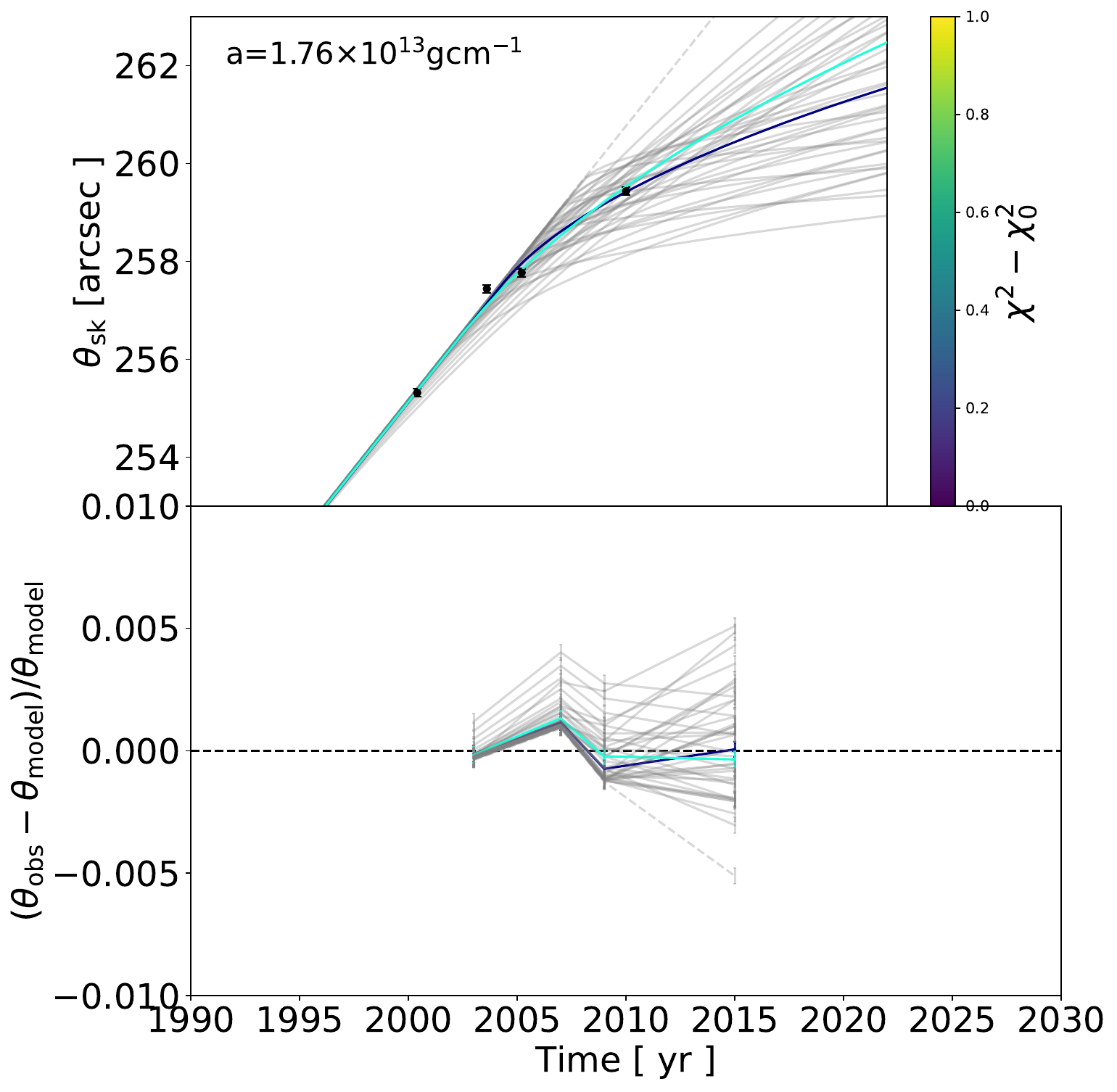}
    \caption{Same as Figure~\ref{fig:a-shift-iso}(b) but for the case of an anisotropic wind. The colorbar is identical to that in Figure ~\ref{fig:a-shift-iso}(b) but now scaled with $\chi_0^2\sim18.3$ for Region 9.}
    \label{fig:a-shift-aniso}
\end{figure}
%\plotone{Re3-1_tchicomp-3-1.pdf}

\begin{figure*}[ht]
    \epsscale{1.15}
    \gridline{\fig{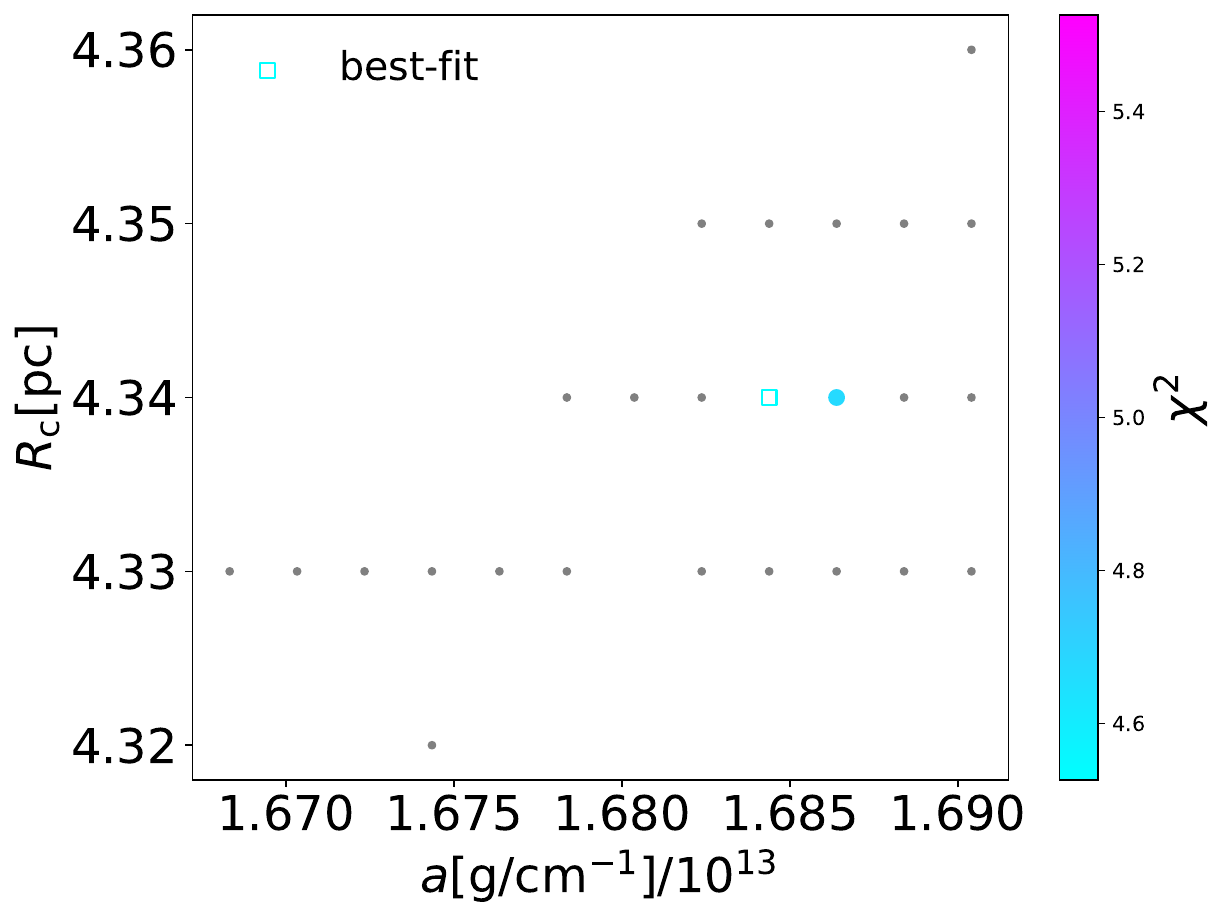}{0.5\textwidth}{(a) Region 1}
    \fig{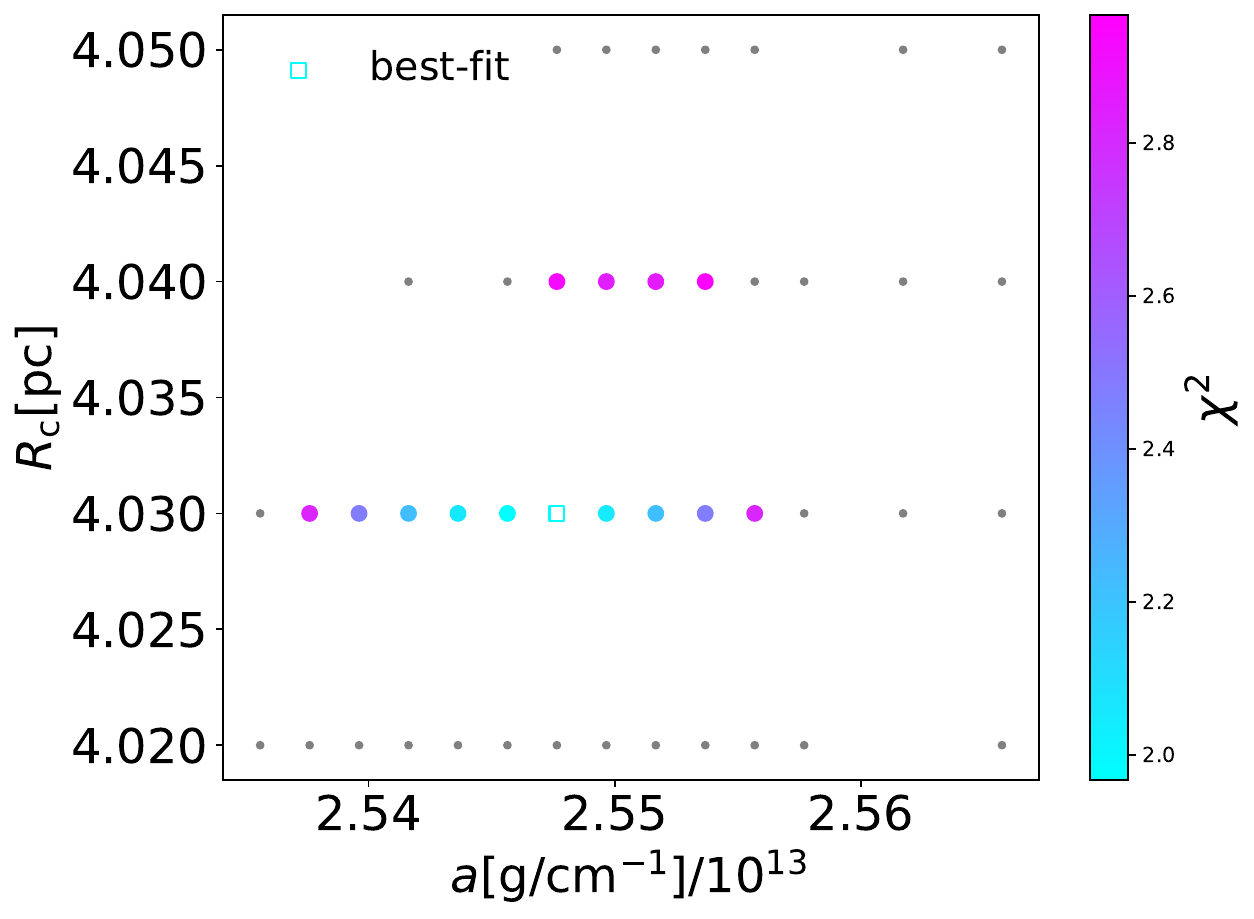}{0.5\textwidth}{(b) Region 6}}
    \caption{The $\chi^2$ value corresponding to each parameter set of $(a,R_\mathrm{c})$ when fixing ($\rho_\mathrm{mc},\Delta r$) to their corresponding best-fit values. Models satisfying the criterion $\chi^2\leq\chi_0^2+1$ are shown by large points colored according to their  $\chi^2$ value, whereas an open square is used for the best-fit model with $\chi^2=\chi_0^2$. The small grey points are for models with $\chi^2>\chi_0 + 1$. 
    %This judgement is lasted until the boundary $\chi^2\gtrless\chi_0^2+1$ is constrained in $(a,R_\mathrm{c})$ plain. 
    The results for Region 1 (a) and Region 6 (b) are shown as examples. The geometric explosion center is assumed for the results shown here.
    %represented as the models with a large error bar and that with a small error bar.
    }
    \label{fig:a-chi}
\end{figure*}

A similar methodology as in Section~\ref{subsubsec:iso} is applied for the anisotropic wind case but now with the wind parameters treated independently for the different regions. As shown in Figure~\ref{fig:a-shift-aniso}, the data for Region 9 can now be well fitted this time using an anisotropic wind model. 

During the parameter survey, we explore $3\times3$ patterns for the cloud density $\rho_\mathrm{mc}$ and the transition length $\Delta r$ for each combination of cloud position $R_\mathrm{c}$ and wind parameter $a$. We pick the best-fit parameter set $(\rho_\mathrm{mc},\Delta r)$ with the minimum $\chi^2$ out of those $3\times3$ models, and repeat the survey on the wind parameter $a$ (except Regions 7 and 8 for which we can use the fixed $a=a_\mathrm{region}$ for the reasons discussed above), and $R_\mathrm{c}$. The result on the best-fit parameter sets in all regions are summarized in Table~\ref{tab:best-geopres}, where the distance $D$ is assumed to be $\sim3.5$~kpc ($\sim3.7$~kpc if the pressure center is assumed) within the range estimated in Section~\ref{subsec:a-d-win}. The result of $\chi^2$ over the explored parameter space 
%(after fixing ($\rho_\mathrm{mc},\Delta r$)) 
is shown in Figure~\ref{fig:a-chi} for Regions 1 and 6 as an example, from which we can see that an effective constraint can be obtained. 
%From this figure, when fixing ($\rho_\mathrm{mc},\Delta r$) we constrain well a parameter set ($a, R_\mathrm{c}$). 

\subsection{Emerging picture of Tycho's CSM}\label{subsec:csm-look}
\begin{figure*}[ht]
    \epsscale{1.15}
    \gridline{\fig{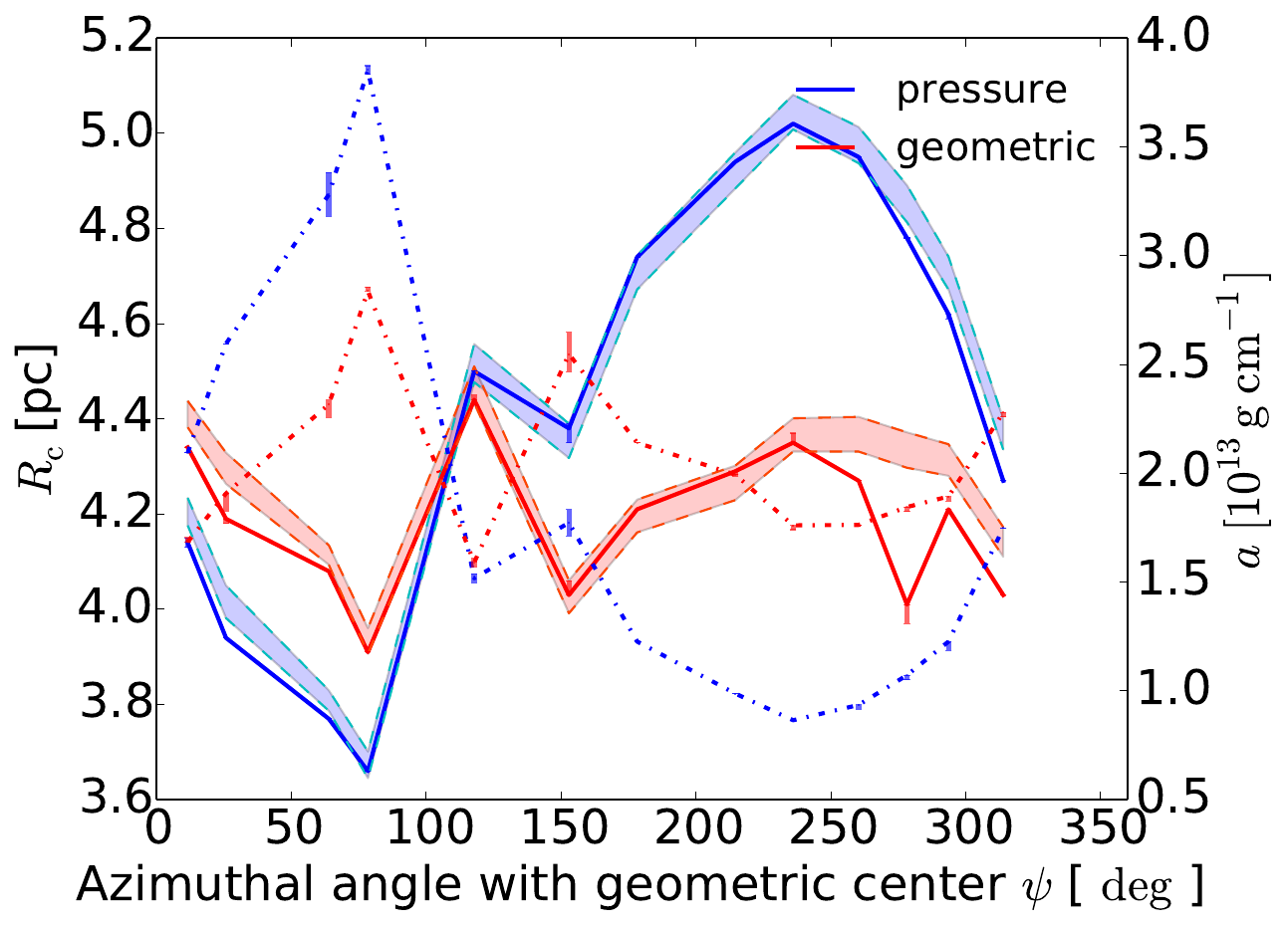}{0.45\textwidth}{(a) Best-fit $a(\psi)$ and $R_\mathrm{c}(\psi)$}
    \fig{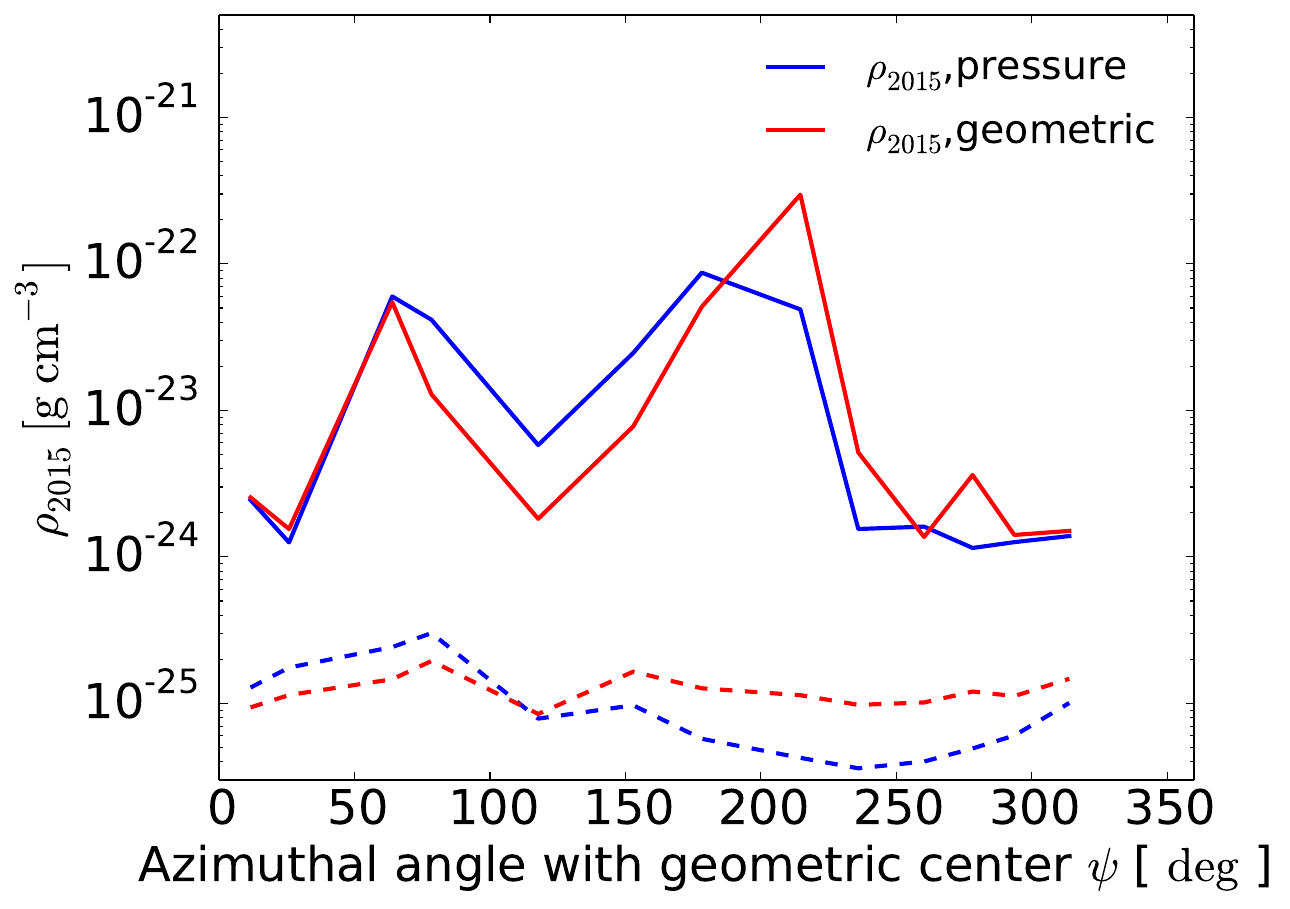}{0.45\textwidth}{(b) Best-fit $\rho(\psi)$}}
    \caption{Best-fit output parameters as a function of azimuthal angle. Panel (a): The red solid line shows the cloud position $R_\mathrm{c}$ and the shaded band shows the radial range between the shock position $R_{2003},R_{2015}$ in the 2003 and 2015 epochs, assuming the geometric explosion center. The blue line and shade are the counterparts assuming the pressure-gradient explosion center. The dashed-dotted lines indicate the corresponding best-fit wind parameter $a$. Panel (b): Estimated gas densities versus azimuthal angle. The solid lines show the estimated ambient gas density at the shock position in 2015. The dashed lines show the density at the position where the free-expanding wind ends (i.e., at the cavity outer boundary). 
    %These parameters is shown only for our best-fit model, no errorbar.
    }
    \label{fig:rmc-phi}
\end{figure*}
%\begin{figure}[p]%[ht]
%    \epsscale{1.15}
%    \plotone{rhomc_15c.pdf}
%    \caption{Estimated densities over azimuthal angles. Blue solid line shows the density where the shock reaches at 2015 using the pressure center (Red solid line is for the geometric center. ), and the blue (and red) dashed lines shows the density at the position where the free-expanding wind ends using the pressure center (and geometric center). These parameters is shown only for our best-fit model, no errorbar.}
%    \label{fig:rhomc-phi}
%\end{figure}

Using the $\chi^2$ values, we find that the best-fit $R_\mathrm{c}$ can either be smaller than $R_{2003}$ or between $R_{2003}$ and $R_{2015}$ depending on the regions, as shown in Figure~\ref{fig:rmc-phi}(a). This trend shows that the shock in some regions had already started experiencing cloud interaction before 2003, and the shock front in the other regions has started penetrating into the dense wall between 2003 and 2015. Indeed, H$\alpha$ filaments have been observed in some regions especially in the east \citep{2010ApJ...715L.146L}, which can result from shock-cloud interaction started earlier on than the other regions. In any case, we find that the FS of Tycho has been propagating in a tenuous wind bubble until only very recently for all azimuthal regions, meaning that the shape of the wind bubble can be well approximated by the current FS position. This makes the discrimination between models with different assumed explosion centers difficult from the point of view of FS dynamics. On the other hand, wind density $a$ has a nearly anti-correlation with $R_{2003}$ and $R_{2015}$ and thus $R_\mathrm{c}$ but this may be attributed to our procedure determining $D$ and $a$ from $R_{2003}$ and $R_{2015}$ which may not be consistent with a real trend.
%note that these discussions show some difference but we do not have enough information yet to discuss off-centered explosion (i.e., The shift between the explosion center and center of CSM or progenitor) because shock-cloud interaction has just begun so recently that we cannot dictate the obvious difference between CSM position and shock position ($R_\mathrm{c}\sim R_{2003},R_{2015}$) which is overall trends despite of which the explosion center is taken. 
%compared with radio contour of Tycho. 

Also, as shown in Figure~\ref{fig:rmc-phi}(b), our best-fit model in each region predicts a steep density gradient from the large gap between $\rho(R_\mathrm{c})$ and $\rho_{2015}\equiv\rho(R_\mathrm{c})+(R_{2015}-R_\mathrm{c})\frac{\partial\rho}{\partial r}$, where $\rho(R_\mathrm{c})$ is the gas density at the outer boundary of the wind cavity and $\rho_{2015}$ is the density of the cloud that the shock is interacting at 2015, respectively. This density structure, which steeply increases outward, does imply an origin of the dense structure from a wind shell swept by a progenitor wind as suggested by \citet{2016ApJ...826...34Z} and \citetalias{2021ApJ...906L...3T} before. The bumpy variation of $\rho_{2015}$ over the azimuth is inconsistent with a more-or-less isotropic and uniform dense medium, but is attributable to the clumpiness of the surrounding cloud. 
%If we assume  we cannot reproduce this deviation which is similar trend in two explosion center assumed.
%One possibility for this relatively large progenitor wind compared by canonical explosion mechanism \citep{1996ApJ...470L..97H,1999ApJ...519..314HachiUme} for SD scenario, is different mass-loss rate over time. This comes from the gap between feature of SN spectrum suggesting normal Ia explosion and large amount of CSM inferred from this work. 

%Table.1
%% Tables should be submitted one per page, so put a \clearpage before
%% each one. maybe automatically clear page if *
%\clearpage
%\begin{turnpage}
\begin{longrotatetable}
\begin{deluxetable*}{l|cccccccccccccc}
\centering
\tablecolumns{14}
\tablewidth{15cm}
\tablecaption{Best-fit model parameters per azimuthal region} 
\tablehead{
Region\tablenotemark{*}&1&2&3&4&5&6&7&8&9&10  &11&12&13
}
\startdata
\hline
Panel A: Geometric center\\
$\dot{M}$ [$10^{-4}\ \mathrm{M_\odot yr^{-1}}$]\tablenotemark{\dag1}&0.839&0.950&1.154&1.413&0.792&1.269&1.066&0.993&0.877&0.878  &0.919&0.611&1.135 \\
%&($_{(<0.001)}^{+0.001}$)&($_{-0.004}^{(<0.001)}$)&($_{-0.003}^{+0.002}$)&($_{(<0.001)}^{+0.001}$)&($_{-0.001}^{+0.001}$)&($_{-0.004}^{+0.005}$)&(-)&(-)&($_{-0.001}^{(<0.001)}$)&(${(<0.001)}$)  &($_{-0.001}^{(<0.001)}$)&($_{-0.001}^{+0.001}$)&($_{(<0.001)}^{+0.001}$) \\
$a$ [$10^{13}\ \mathrm{g\ cm^{-1}}$]\tablenotemark{\dag1}&1.68&1.91&2.32&2.84&1.59&2.55&2.14&1.99&1.76&1.76  &1.85&1.89&2.28\\
$R_\mathrm{c}$ [pc]\tablenotemark{\dag2}&4.34&4.19&4.08&3.91&4.44&4.03&4.21&4.29&4.35&4.27  &4.01&4.21&4.03 \\
%&($<0.01$)&($_{-0.01}$)&($<0.01$)&($<0.01$)&($^{+0.01}$)&($^{+0.03}$)&($<0.01$)&$_{-0.01}$&($^{+0.02}$)&($<0.01$)  &($^{+0.04}$)&($<0.01$)&($<0.01$) \\
$\frac{\partial\rho}{\partial r}$ [$10^{-23}$gcm$^{-3}$pc$^{-1}$]\tablenotemark{\dag3}&2.48&0.99&100&25&2.5&25&250&2500&10&0.99  &0.99&0.99&0.99 \\
$R_{2015}-R_\mathrm{c}$ [pc]\tablenotemark{\dag4}&0.098&0.14&0.055&0.050&0.070&0.031&0.030&0.012&0.051&0.13&  0.36& 0.14&0.14 \\
$\rho_{2015}$ [$10^{-24}$gcm$^{-3}$]\tablenotemark{\dag5}&2.54&1.55&54.8&12.9&1.82&7.78&51.0&297&5.16&1.37  &3.62&1.41&1.50 \\ \\ \hline
Panel B: Pressure center\\
$\dot{M}$ [$10^{-4}\ \mathrm{M_\odot yr^{-1}}$]\tablenotemark{\dag1}&1.043&1.293&1.634&1.919&0.756&0.883&0.611&0.492&0.431&0.466 
 &0.534&0.611&0.871 \\
%&($<0.001$)&($_{(<0.001)}^{+0.001}$)&($_{-0.005}^{+0.005}$)&($_{-0.001}^{+0.001}$)&($_{-0.001}^{+0.001}$)&($_{-0.003}^{+0.003}$)&(-)&(-)&($<0.001$)&($_{-0.001}^{(<0.001)}$)  &($_{-0.001}^{+0.001}$)&($_{-0.002}^{+0.001}$)&($<0.001$) \\
$a$ [$10^{13}\ \mathrm{g\ cm^{-1}}$]\tablenotemark{\dag1}&2.09&2.60&3.28&3.85&1.52&1.77&1.23&0.988&0.865&0.936&1.07&1.23&1.75\\
$R_\mathrm{c}$ [pc]\tablenotemark{\dag2}&4.14&3.94&3.77&3.66&4.50&4.38&4.74&4.94&5.02&4.95  &4.78&4.62&4.27 \\
%&($_{-0.01}^{+0.01}$)&($<0.01$)&($<0.01$)&($<0.01$)&($<0.01$)&($_{-0.03}^{+0.01}$)&($<0.01$)&($<0.01$)&($<0.01$)&($<0.01$)  &($<0.01$)&($_{-0.01}$)&($<0.01$) \\
$\frac{\partial\rho}{\partial r}$ [$10^{-23}$gcm$^{-3}$pc$^{-1}$]\tablenotemark{\dag3}&2.5&0.98&100&100&10&250&2500&250&25&250  &1.0&100&0.99 \\
$R_{2015}-R_\mathrm{c}$ [pc]\tablenotemark{\dag4}&0.094&0.11&0.60&0.041&0.057&0.0098&0.0035&0.020&0.061&0.063&  0.11&0.12&0.13 \\
$\rho_{2015}$ [$10^{-24}$gcm$^{-3}$]\tablenotemark{\dag5}&2.44&1.26&59.9&41.5&5.80&24.7&87.1&48.9&1.55&158  &1.15&121&11.39 \\
\enddata
\tablenotetext{*}{Regions are numbered according to the position (azimuthal) angles following the definition in \citetalias{2021ApJ...906L...3T}. The distance $D$ is assumed to be $\sim3.5$~kpc and $\sim3.7$~kpc when the geometric and pressure center are used.}
\tablenotetext{\dag1}{\ \ Wind density profile is assumed to be $\rho(r)=a r^{-2}$, where $a=\dot{M}/(4\pi V_\mathrm{w})$ and wind velocity is fixed at $V_\mathrm{w}=250$~km/s. }%Rough errorbar of $\Delta\chi^2=1$ with changing parameter $R_\mathrm{c}$ are in parentheses.}
\tablenotetext{\dag2}{\ \ Best-fit radius of the outer boundary of the free-expanding wind.}% Rough errorbar of $\chi^2-\chi_0^2=1$ where $\chi_0^2$ is the minimum of $\chi^2$ with changing parameter $a$ in parentheses.}
\tablenotetext{\dag3}{\ \ Average density gradient between the outer boundary of the free-expanding wind and the dense cloud.}
\tablenotetext{\dag4}{\ \ Radial separation of the FS in year 2015 from the cavity boundary.}
\tablenotetext{\dag5}{\ \ Density at $r=R_{2015}$.}

\label{tab:best-geopres}
\end{deluxetable*}
\end{longrotatetable}

\section{Discussion}\label{sec:disc}

\subsection{On the inferred wind properties}\label{subsec:disc-wind-explain}
%\begin{figure}[ht]
    %\epsscale{1.15}
    %\plotone{Mwind_1-1.pdf}
    %\caption{Estimated wind mass $4\pi aR_\mathrm{region}$ over azimuthal angles, which is the total wind mass assuming the same wind density as $a_\mathrm{region}$ at the region over all azimuthal angles, is shown in blue solid line for the pressure center (red line for the geometric center). The horizontal line in the same color shows each estimated the symmetry over $\phi$ (see the text). These parameters is shown only for our best-fit model, no errorbar.}
    %\label{fig:mwind-phi}
%\end{figure}
From our calculations above, we have obtained order estimates for the wind properties in the cavity surrounding Tycho, with $a\sim2\times10^{13}\ \mathrm{g\ cm^{-1}}$ where $\rho(r)=ar^{-2}$ is assumed. %First, the variation of $a$ over position angle can be connected to the geometry of the wind. If we consider accretion wind like Hachisu wind \citep{1996ApJ...470L..97H,1999ApJ...519..314HachiUme}, the disk-like structure for mass-loss is expected. 
The total mass in the free-expanding wind zone can then be expressed in polar coordinates $(r, \psi, \phi)$ 
\begin{eqnarray}
    M_\mathrm{wind}&=&\\ \nonumber
    \int_0^{R_\mathrm{c,Region}}&r&^2dr\int_{-1}^{1}d(\cos(\psi-\delta\psi))\int_0^{2\pi}d\phi\ ar^{-2}\\
    \sim1.635&\pm&0.085~\mathrm{M_\odot}(1.18\pm0.08~\mathrm{M_\odot}\mathrm{;\ pressure\ center}), \nonumber
\end{eqnarray}
where $\psi$ is the azimuthal angle and $\phi$ is the polar angle. Here we have assumed that the wind is symmetric over $\phi$ such that the sky-projected wind bubble is symmetric along $\psi=\delta\psi$. For $\delta\psi$, we use $\delta\psi=78.4^\circ$ and $260.3^\circ$ roughly corresponding to Region 3 and Region 10 where the maximum and minimum of $aR_\mathrm{c,Region}$ are located. The estimated $M_\mathrm{wind}$ suggests that the pre-SN mass-loss is about one to a few $\gtrsim M_\odot$\footnote{We note that this estimate can vary to some extent if Fe K$\beta$ emission is used as the RS locator instead of K$\alpha$, as we have mentioned above.}. %%If the mass is a few~$M_\odot$ the SN explosion will be Ia-CSM type. 
%a lower limit because we use the value for RS angular size being  which is somewhat larger than . }. 
The derived CSM mass far exceeds the mass budget available in the DD scenario. On the other hand, this is consistent with the SD scenario, where the mass budget for the CSM can be attributed to the companion star whose initial mass is $\sim2-4\ M_\odot$\citep[][]{1999ApJ...522..487H}. 
DD scenario is difficult to explain the CSM property because such a large mass-loss can hardly occur during the pre-SN activities. %It is possible that the assumed wind velocity is too large and can be a factor of 10 slower. 

The mass-loss rate could however place stronger constraints on the underlying scenarios, even for the SD scenario. With $V_{\rm w} = 250$ km s$^{-1}$, the pre-SN activity must have operated in the final $\sim 10000$ yrs with the mass-loss rate of $\sim 10^{-4} M_\odot$ yr$^{-1}$. However, the mass `accretion' rate onto the progenitor WD is believed to be limited to $< 10^{-6} M_\odot$ in the canonical SD scenario toward a near-Chandrasekhar WD \citep{1984ApJ...277..791N}. Therefore, only $\sim 0.01 M_\odot$ of the mass was accumulated during this final phase, which unlikely leads to the formation of a near-Chandrasekhar WD. This may be remedied if $V_{\rm w}$ is lower, e.g., $\sim 25$ km s$^{-1}$, for which the mass accretion can exceeds $\sim 0.1 M_\odot$; it will probably require a slow wind associated with an extended donor, rather than the WD wind. In this case, the extended envelope might have been stripped away by the SN shock interaction \citep[e.g.,][]{2000ApJS..128..615M,2008AA...489..943P,2012AA...548A...2L,2012ApJ...750..151P}, and thus may leave only a very faint surviving companion which might be consistent with non-detection of a surviving companion star \citep{2004Natur.431.1069R,2009ApJ...701.1665K,2015ApJ...809..183X}. The above constraint could also be overcome in the double detonation scenario, i.e., an explosion initiated by the runaway triple-alpha reactions on the accumulated He layer of the progenitor WD \citep{1982ApJ...253..798N,1990ApJ...354L..53L,2009ApJ...699.1365S} -- this scenario does not require substantial increase of the WD mass toward the explosion. 

In view of the combination of the large mass budget and the high accretion rate, another appealing scenario is the core-degenerate scenario and its variants \citep[e.g.,][]{1974ApJ...188..149S,2015MNRAS.450.1333S}. This scenario may explain the large CSM mass by the common envelope interaction, and the ignition of the thermonuclear runaway by a rapid He accretion during the core merger of a WD and He-rich companion \citep[e.g.,][]{2020Sci...367..415J}. We however note that such scenario has not been investigated in details (not to the level the SD and DD scenarios have been investigated), and thus its applicability to Tycho is still quite speculative. 
%Since, e.g., an accretion wind around a white dwarf such as that in the Hachisu model \citep{1996ApJ...470L..97H,1999ApJ...519..314HachiUme} is coming from the companion star of which the mass budget is around $\lesssim3M_\odot$, this estimate $\gtrsim 1M_\odot$ is only marginally consistent with the scenario of an accretion wind. 
%If this mass is even larger, it will become inconsistent with the light echo \citep{2008Natur.456..617K} and the X-ray thermal spectra \citep{2006ApJ...645.1373B}, which identify Tycho as a normal Ia SNR. 

%by One possibility for this relatively large mass-loss rate 
%The inferred mass-loss rate can then be suppressed from $\dot{M}\sim10^{-4}\ \mathrm{M_\odot/yr}\ (V_w=250\ \mathrm{km/s})$ to $\dot{M}\sim10^{-5}\ \mathrm{M_\odot/yr}\ (V_w=25\ \mathrm{km/s})$.%but barely explaining a stable accretion onto the WD in a SD scenario. 
%Another possibility is the core-degenerate scenario with a sub-Chandrasekhar mass progenitor. We note however that these discussions are based on our 1-D simulations, and multi-dimensional effects may lead to important modifications, which is a topics to pursue in our future work. 
%Hachisu wind \citep{1996ApJ...470L..97H,1999ApJ...519..314HachiUme} for SD scenario, is different mass-loss rate over time. This comes from the gap between feature of SN spectrum suggesting normal Ia explosion and large amount of CSM inferred from this work. Another possibility is after common envelope forms moderate explosion occurs. 

\subsection{Comparison with previous studies}\label{subsec:disc-prev}
\subsubsection{The size of wind bubble}\label{subsec:disc-wind-time}

There are previous theoretical studies which considered wind-like environment models. This model is designed to reproduce the dense shocked ejecta shell required by the observed high ionization degree \citep{2006ApJ...645.1373B} in the inner region, and the low density suggested from the expansion parameter \citep{2010ApJ...709.1387K,1997ApJ...491..816R} simultaneously. %From the condition that the shock is in a free-expanding wind and the ejecta shell is dense, according to 
One of such studies, \citet{2013MNRAS.435.1659C} \citep[see also][]{Slane2014}, used more compact wind bubble less than 1.8 pc, which is much less than the model in the present study. %due to a shorter wind duration time assumed. 
%showing a lesser influence of the wind on the long-term shock evolution

Now that \citet{2016ApJ...826...34Z} have shown that the shock is interacting with a dense gas shell, and \citetalias{2021ApJ...906L...3T} and recent Doppler motion analysis \citep{Kasuga-tycho} are suggesting that the shock-cloud interaction has just begun around 2007 on average, a more consistent picture seems to point to a wind-shell radius nearly equal to the current shock radius, as suggested by the present work. 
The $r^{-2}$ density profile assumed in this work in the wind cavity extending up to nearly the current FS radius can explain the dense shocked ejecta and a low density surrounding environment ($\sim 10^{-25}\ \mathrm{g cm^{-3}}$ at a radius of $\sim$ 3 pc) to reconcile with the observations. 

\subsubsection{Anisotropic wind or ejecta?}\label{subsec:disc-not-density}

\citet{2022ApJ...937..121M} explained the azimuthal variation of the shock evolution using the azimuthal variation of the explosion properties by putting a larger kinetic energy into certain angles, e.g., the southeast region. While this time we have attributed the azimuthal variation of the shock proper motion mainly to the azimuthal variation of the wind density and the radial position of the dense cloud, a combined effect with an asymmetric explosion remains possible. However, recent 3-D hydro simulations of Type Ia SNRs like Tycho from the SN phase have indicated that the SNR morphology will lose memory of its original asymmetric SN properties within a time scale of $\sim 100$ years after explosion \citep[e.g.,][]{Ferrand_2019}. %there seems to remain the variation of the shocked ejecta. 

\section{Summary}\label{sec:sum}
Among Type Ia SNRs, Tycho has long been considered to be a prototypical remnant from the viewpoints of morphology, ambient environment, X-ray spectrum and light echo. However, recent results of Tycho’s radio and X-ray observations \citep{2010ApJ...709.1387K,2016ApJ...823L..32W} have shown azimuthal variability of the shock motion, and CO observation \citep{2016ApJ...826...34Z} has suggested that Tycho is surrounded by a cloud with a cavity swept up by a past wind-like activity, which supported a SD scenario. Moreover, reanalysis of Tycho's Chandra data has shown that since around 2007 Tycho's shock has been experiencing a substantial deceleration, which infers a recent interaction with molecular cloud supporting a picture of a cavity-wall environment. Such a non-uniform environment has strong implications on the nature of Tycho's progenitor system. 

We have performed 1-D hydro simulations to model the latest multi-epoch proper motion measurements of Tycho's shock. 
%and fit with the observed  data. with the distance $D_{\rho_\mathrm{c}}$ corresponded to cavity density $\rho_\mathrm{c}$. 
We have tested two scenarios for the density structure in the cavity surrounded by a dense cloud, i.e., a uniform-density medium and a wind-like cavity. In the uniform cavity case, all models fail to reproduce the large angular velocity in regions like Regions 7 and 8 even by invoking unrealistically low ambient density as $2.0\times10^{-27} \mathrm{gcm^{-3}}$, leading us to consider a wind-like environment. In the wind-like cavity case, an anisotropic wind can provide a satisfactory fit to all regions. 
%the parameter obtained in the same procedure can reproduce more likely the data compared to uniform cavity case. 
The best-fit parameter set in each region indicates a disk-like instead of a spherical isotropic mass-loss activity, with the wind density to be $a\sim2\times10^{13}\ \mathrm{g\ cm^{-1}}$ where $\rho(r)=ar^{-2}$. This can be dictated to the mass-loss rate estimated to be $\sim1\times10^{-5}\ \mathrm{M_\odot\ yr^{-1}}$ for a wind velocity of 25~$\mathrm{km\ s^{-1}}$, which may lead to the formation of a near-Chandrasekhar explosion. The total mass-loss is estimated to be one to a few~$M_\odot$, which is within the scope of the SD scenario. %is higher than expected from a typical accretion wind model for white dwarfs. 
%This issue may be lifted if a lower wind velocity $V_\mathrm{w}\sim25\ \mathrm{km/s}$ is invoked, so that our results are consistent with a SD scenario. %One possibility for this relatively large progenitor wind compared by canonical explosion mechanism \citep{1996ApJ...470L..97H,1999ApJ...519..314HachiUme} for SD scenario, is different mass-loss rate over time. This comes from the gap between feature of SN spectrum suggesting normal Ia explosion and large amount of CSM inferred from this work. 
Finally, our results have revealed an azimuthal variation of the cloud density, which can be explained by the clumpiness of the cloud.

\begin{acknowledgements}
We thank Professor Takashi Hosokawa at the department of physics of Kyoto University for fruitful discussion on parameters of cloud. This work is supported by JSPS grant Nos. JP19K03913 (S.H.L.), JP19H01936 (T.T.), JP21H04493 (T.T.), and JP20H00174 (K.M.). R.K. acknowledges support by Science Faculty Scholarship from the Kyoto University Foundation, and JST, the establishment of university fellowships towards the creation of science technology innovation, Grant Number JPMJFS2123. S.H.L. acknowledges support by the World Premier International Research Center Initiative (WPI), MEXT, Japan. 
\end{acknowledgements}

%\section{table}

\bibliography{reference}{}
\bibliographystyle{aasjournal}

\end{document}